\documentclass[aps,prx,onecolumn,citeautoscript,superscriptaddress,nofootinbib,
eqsecnum,preprintnumbers]{revtex4}  
\usepackage{amsmath,amssymb,bm} 
\usepackage{graphicx}  
\usepackage{color}
\usepackage[papersize={8.5in,11in}]{geometry}
\usepackage[colorlinks=true]{hyperref}  
\hypersetup{
    unicode=false,          % non-Latin characters 
    pdftoolbar=true,        % show Acrobat
    pdfmenubar=true,        % show Acrobat 
    pdffitwindow=false,     % window fit to page when opened
    pdfstartview={FitH},    % fits the width of the page to the window
    pdftitle={Magnetotransport in a model of a disordered strange metal}, % title
    pdfauthor={Aavishkar A. Patel, John McGreevy, Daniel P. Arovas and Subir Sachdev},     % author
    pdfsubject={},   % subject of the document
    pdfcreator={},   % creator of the document
    pdfproducer={}, % producer of the document
    pdfkeywords={} {} {}, % list of keywords
    pdfnewwindow=true,      % links in new window
    colorlinks=true,       % false: boxed links; true: colored links
    linkcolor=magenta,  % color of internal links (change box color with linkbordercolor)
    citecolor=blue,        % color of links to bibliography
    filecolor=magenta,      % color of file links
    urlcolor=blue           % color of external links
} 

\usepackage{amsfonts}
\usepackage{upgreek}
\usepackage{slashed}
\usepackage{latexsym}

\newcommand{\beq}{\begin{equation}}
\newcommand{\eeq}{\end{equation}}
\newcommand{\nn}{\nonumber \\}

\def \dg{\dagger}

\newcommand{\change}[1]{{\color{black} #1}}
\newcommand{\rchange}[1]{{\color{black} #1}}
\date{\today}

\preprint{NSF-ITP-17-154}

\begin{document}

\title{Magnetotransport in a model of a disordered strange metal}

\author{Aavishkar A. Patel}
\affiliation{Department of Physics, Harvard University, Cambridge MA 02138, USA}
\affiliation{Kavli Institute for Theoretical Physics, University of California, Santa Barbara CA 93106-4030, USA}

\author{John McGreevy}
\affiliation{Department of Physics, University of California at San Diego, La Jolla, CA 92093, USA}

\author{Daniel P. Arovas}
\affiliation{Department of Physics, University of California at San Diego, La Jolla, CA 92093, USA}

\author{Subir Sachdev}
\affiliation{Department of Physics, Harvard University, Cambridge MA 02138, USA}
\affiliation{Perimeter Institute for Theoretical Physics, Waterloo, Ontario, Canada N2L 2Y5}
\affiliation{Department of Physics, Stanford University, Stanford CA 94305, USA}

\begin{abstract}
\rchange{Despite much theoretical effort, there is no complete theory of the `strange' metal state of the high temperature superconductors, and its linear-in-temperature, $T$, resistivity. Recent experiments showing an unexpected linear-in-field, $B$, magnetoresistivity have deepened the puzzle. We propose a simple model of itinerant electrons, interacting via random couplings with electrons localized on a lattice of quantum `dots' or `islands'. This model is solvable in a particular large-$N$ limit, and can reproduce observed behavior. The key feature of our model is that the electrons in each quantum dot are described by a Sachdev-Ye-Kitaev model describing electrons without quasiparticle excitations.} For a particular choice of the interaction \rchange{between the itinerant and localized electrons}, this model realizes a controlled description of a diffusive marginal-Fermi liquid (MFL) without momentum conservation, which has a linear-in-$T$ resistivity and a $T\ln T$ specific heat as $T \rightarrow 0$. By tuning the strength of this interaction relative to the bandwidth of the itinerant electrons, we can additionally obtain a finite-$T$ crossover to a fully incoherent regime that also has a linear-in-$T$ resistivity. We describe the magnetotransport properties of this model, \change{and show that the MFL regime has conductivities which scale as a function of $B/T$; however, the magnetoresistance saturates at large $B$.} We then consider a macroscopically disordered sample with domains of such MFLs with varying densities of electrons and islands. Using an effective-medium approximation, we obtain a macroscopic electrical resistance that scales linearly in the magnetic field $B$ applied perpendicular to the plane of the sample, at large $B$. The resistance also scales linearly in $T$ at small $B$, and as $T f(B/T)$ at intermediate $B$. We consider implications for recent experiments reporting linear transverse magnetoresistance in the strange metal phases of the pnictides and cuprates.
\end{abstract}

\maketitle

\section{Introduction}
\label{intro}

Essentially all correlated electron high temperature superconductors display an anomalous metallic state at temperatures above the superconducting critical
temperature at optimal doping \cite{Keimer15,Matsuda2010,SK2011}. This metallic state has a `strange' linearly-increasing dependence of the resistivity, $\rho$, on temperature, $T$;
it can also exhibit bad metal behavior with a resistivity much larger than the quantum unit $\rho \gg h/e^2$ (in two spatial dimensions) \cite{Emery95}.
More recently, strange metals have also been demonstrated to have a remarkable 
linear-in-$B$ magnetoresistance, with the crossover between the linear-in-$T$ and linear-in-$B$ behavior occurring at $\mu_BB \sim k_B T$ \cite{Hayes2016,Giraldo2017}.

\change{
This paper will present a model of a strange metal which exhibits the above 
linear-in-$T$ {\it and\/} linear-in-$B$ behavior. The model builds on a lattice array of
quantum `dots' or `islands', each of which is described by a Sachdev-Ye-Kitaev (SYK) model of fermions with random all-to-all interactions \cite{SY93,kitaev2015talk}. 
The SYK models are 0+1 dimensional quantum theories which exhibit a `local criticality'. 
They have drawn a great
deal of interest for a variety of reasons:
\begin{itemize}
\item The SYK models are the simplest solvable models without quasiparticle excitations. They can also be used as fully quantum building blocks
for theories of strange metals in non-zero spatial dimensions \cite{PG98,Balents2017}.
\item The SYK models exhibit many-body chaos \cite{kitaev2015talk,Maldacena2016}, and saturate the lower bound on the Lyapunov time to reach chaos \cite{Maldacena2016a}. So they are ``the most chaotic'' quantum many-body systems. The presence of maximal chaos is linked to the absence of quasiparticle excitations,  and the proposed \cite{ssbook} lower bound of order $\hbar /(k_B T)$  on a `dephasing time'. 
It is important to note here that the co-existence of many-body chaos and solvability is quite remarkable: essentially all other solvable models
({\em e.g.\/} integrable lattice models in one dimension) do not exhibit many-body chaos.
\item Related to their chaos, the SYK models exhibit \cite{Sonner17} eigenstate thermalization (ETH) \cite{Deutsch91,Srednicki94}, and yet many aspects are exactly solvable.
\item The SYK models are dual to gravitational theories in $1+1$ dimensions which have a black hole horizon. The connection between the SYK models and black holes with a near-horizon AdS$_2$ geometry was proposed in Refs.~\cite{SS10,SS10b}, and made much sharper in Refs.~\cite{kitaev2015talk,nearlyads2,kitaev2017}. This connection has been used to examine aspects of the black hole information problem \cite{Maldacena2017}. 
\end{itemize}
}

More specifically, 
a single SYK site is a 0+1 dimensional non-Fermi liquid in which the imaginary-time ($\tau$) fermion Green's function has the low $T$ `conformal' form \cite{SY93,PG98,Faulkner09,Sachdev2015}
\beq
G(\tau) \sim \left( \frac{T}{\sin(\pi T \tau)} \right)^{1/2} e^{-2 \pi \mathcal{E} T \tau}\,, \quad 0 < \tau < 1/T \,,
\label{Glocal}
\eeq
where $\mathcal{E}$ is a parameter controlling the particle-hole asymmetry. 
\change{
In frequency space, this correlator is $G(\omega) \sim 1/\sqrt{\omega}$ for $\omega \gg T$, and this implies non-Fermi liquid behavior.}
\rchange{A Fermi liquid has the exponent 1/2 in Eq.~(\ref{Glocal}) replaced by unity, and a constant density of states with $G(\omega)$ frequency independent.
The Green's function in Eq.~(\ref{Glocal})} implies \cite{SY93} a `marginal' \cite{Varma89} susceptibility, $\chi$,  with a real part which diverges logarithmically
with vanishing frequency ($\omega$) or $T$. Specifically, in the all-to-all limit of the SYK model, vertex corrections
are sub-dominant, and \rchange{Fourier transform of} $\chi (\tau) = - G(\tau) G(-\tau)$ leads to the spectral density
\beq
\mbox{Im}\, \chi (\omega) \sim \tanh \left( \frac{\omega}{2 T} \right)\,,
\label{chitanh}
\eeq
whose Hilbert transform leads to the noted logarithmic divergence. \rchange{In contrast, a Fermi liquid has $\mbox{Im}\, \chi (\omega) \sim \omega$.} The
form in Eq.~(\ref{chitanh}) is consistent with recent electron scattering observations \cite{Abbamonte17}. 
A linear-in-$T$ resistivity now follows upon considering itinerant fermions scattering off such a local susceptibility, and the itinerant fermions realize a marginal Fermi liquid (MFL) with a $\omega \ln \omega$
self energy \cite{Varma89,SY93,SS10,Faulkner2013}.

\change{
We now review previous approaches to building a finite-dimensional non-Fermi liquid from the $0+1$ dimensional SYK model.
An early model} for a bulk strange metal in finite spatial dimensions 
was provided by Parcollet and Georges \cite{PG98}.
They considered a doped Mott insulator described by a random $t$-$J$ model at hole density $\delta$,
where $t$ is the root-mean-square (r.m.s.) electron hopping, and $J$ is the r.m.s. exchange interaction. 
At low doping with $\delta t \ll J$, they found strange metal behavior in the intermediate $T$
regime $E_c < T < J$, where the coherence energy $E_c = (\delta t)^2/J$. 
\change{ In this intermediate energy range, they found that the electron Green's function had the local form of the SYK model in 
Eq.~(\ref{Glocal}). Moreover, this metal had `bad metal' resistivity with $\rho \sim (h/e^2) (T/E_c) \gg (h/e^2)$. We will refer to
such a strange metal as an `incoherent metal' (IM). This IM is to be contrasted from a MFL, which we will describe below; the MFL does not appear
in the model of Parcollet and Georges.}

Another finite-dimensional model of an IM appeared in the recent work of Song {\it et al.} \cite{Balents2017}.
They considered a lattice of SYK sites, with r.m.s. on-site interaction $U$, and r.m.s. inter-site hopping $t$. 
\change{Each site was a quantum island with $N$ orbitals, and had random on-site interactions with typical magnitude $U$.
Electrons were allowed to hop between nearest-neighbor states, with a random matrix element of magnitude $t$.
Although this is a model with strong interactions, the remarkable fact is that the random nature of the interactions renders
it exactly solvable.
As in Ref.~\onlinecite{PG98}, Song {\it et al.}} found an IM in the intermediate regime $E_c < T < U$, with a local
electron Green's function as in  Eq.~(\ref{Glocal}), and a bad metal resistivity $\rho \sim (h/e^2) (T/E_c)$.
Their coherence scale was $E_c = t^2/U$. (This lattice SYK model should be contrasted from earlier studies \cite{Gu2017,Sachdev2017},
which only had fermion interaction terms between neighboring SYK sites: the latter models realize disordered metallic
states without quasiparticle excitations as $T \rightarrow 0$, but have a $T$-independent resistivity.)

\change{Although these models \cite{PG98,Balents2017} reproduce bad metal resistivity, we will show here that they are unable to describe the experimentally observed large magnetoresistance
noted earlier \cite{Hayes2016,Giraldo2017}. The random nature of the hopping between the sites, 
and the associated absence of a Fermi surface, results in negligible magnetoresistance. Significant orbital magnetoresistance only appears
in models which have fermions with non-random hopping and a well-defined Fermi surface. Note that the existence of a Fermi surface does not
directly imply the presence of well-defined quasiparticles: it is possible to have a sharp Fermi surface in momentum space (where the inverse
fermion Green's function vanishes) while the quasiparticle spectral function is broad in frequency space.}
 
\begin{figure}
\begin{center}
\includegraphics[height=2.1in]{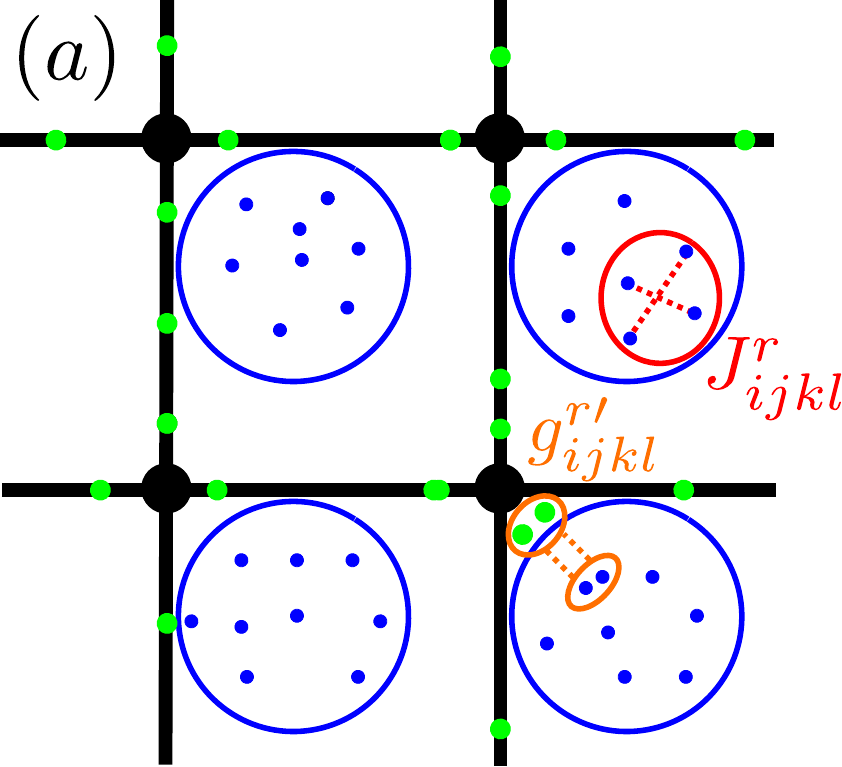} ~~~~~ \includegraphics[height=2.1in]{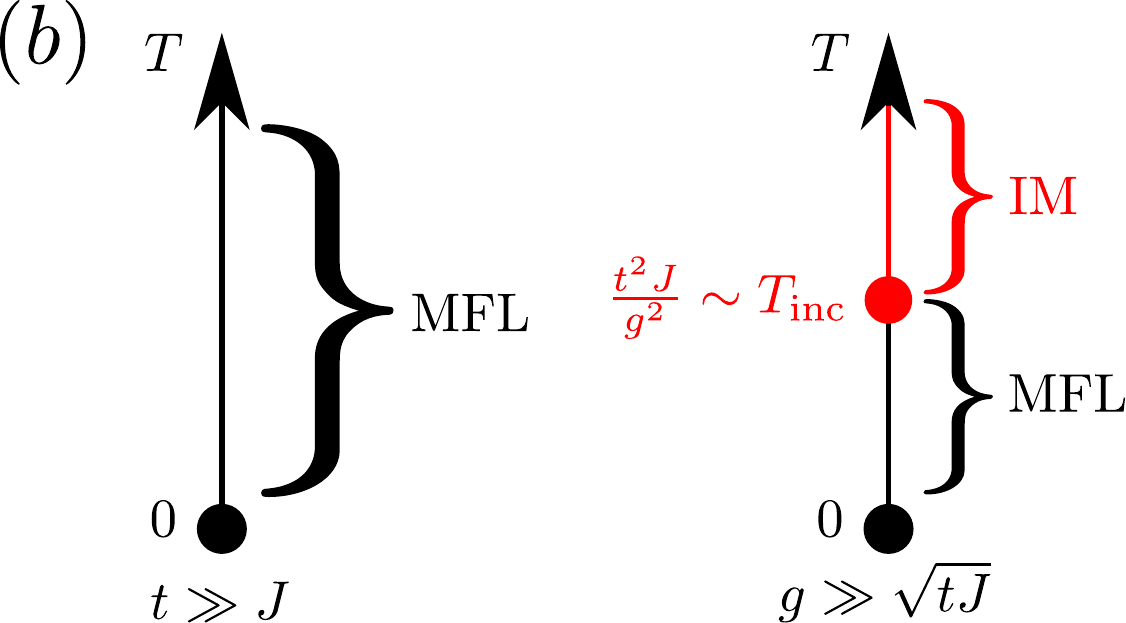}
\end{center}
\caption{(a) A cartoon of our microscopic model. \rchange{Itinerant} conduction electrons (green) hop around on a lattice (black). At each lattice site, they interact locally and randomly with SYK \rchange{quantum dots} (blue) through an interaction (orange) that independently conserves the numbers of conduction and island electrons. (b) Finite-temperature regimes of the model. When the conduction electron bandwidth is large enough, it realizes a disordered marginal-Fermi liquid (MFL) for the conduction electrons for all temperatures $T\ll J$ (Sec.~\ref{infiniband}). For a finite bandwidth, there can be a finite-temperature crossover to an `incoherent metal' (IM), in which all notion of electron momentum is lost, if the coupling $g$ is large enough (Sec.~\ref{dialup}). Note that we always have $J\gg T$ and $J\gtrsim g$.}
\label{Modelfig}
\end{figure}
\change{With the aim of obtaining a well-defined Fermi surface of itinerant electrons, in this paper we 
consider a lattice of SYK islands coupled to a separate band of itinerant \rchange{conduction} electrons \rchange{as illustrated in Fig.~\ref{Modelfig}}.}
Our model is in the spirit of effective Kondo lattice models which have been proposed as models of the physics of the disordered, 
single-band Hubbard model \cite{MSB89,BF92,senthilLM}.
Other two band models of itinerant electrons coupled to SYK excitations have been considered in Refs.~\onlinecite{BGG01,McGreevy2017}. Our model exhibits
MFL behavior as $T \rightarrow 0$, with a linear-in-$T$ resistivity, and a $T \ln T$ specific heat. For an appropriate range of parameters, there is a crossover at higher $T$ to an IM regime, also with a linear-in-$T$ resistivity. The itinerant electrons have
a {\it non\/}-random hopping $t$, the SYK sites have a random interaction with r.m.s. strength $J$, and these
two sub-systems interact with a random Kondo-like exchange of r.m.s. strength $g$: see Fig.~\ref{Modelfig}a for
a schematic illustration. Fig.~\ref{Modelfig}b illustrates the regimes of MFL and IM behavior in our model.
\change{In the MFL regime, our model exhibits a well-defined Fermi surface, albeit of damped quasiparticles.}

The magnetotransport properties of this model will be a significant focus of our analysis. 
\change{We will show that the MFL regime with a Fermi surface indeed has a sizeable magnetoresistance, with 
characteristics in accord with observations.}
We find that the longitudinal and Hall conductivities, \change{of the MFL regime}, can be written as scaling functions of 
$B/T$, as shown in Eq.~(\ref{eq:Bscale}). In contrast, the $B$ dependence is much less singular in the IM regime.
\change{Although a $B/T$ scaling is obtained in the MFL in 
this computation, the magnetoresistance does not increase linearly with $B$,
and instead saturates at large $B$. To obtain a non-saturating magnetoresistance we} 
consider a macroscopically disordered sample with domains of MFLs with varying electron densities;
employing earlier work on classical electrical transport in inhomogeneous ohmic conductors \cite{Dykhne1971,Stroud1975,Parish2003,Parish2005,Guttal2005,Song2015,Ramakrishnan2017}, we obtain 
the observed linear-in-$B$ magnetoresistance with a crossover scale at $B \sim T$.

This paper is organized as follows: In Sec.~\ref{basemodel}, we introduce our basic microscopic model of a disordered MFL, and determine its single-electron properties and finite-temperature crossovers in Sec.~\ref{mflandim}. In Sec.~\ref{Transport}, we solve for transport and magnetotransport properties of this basic model exactly in various analytically-tractable regimes. In Sec.~\ref{EMARRN}, we introduce the effective-medium approximation and apply it to a macroscopically disordered sample containing domains of the basic model, obtaining analytical results for the global magnetotransport properties for certain simplified considerations of macroscopic disorder. We summarize our results and place them in the context of recent experiments in Sec.~\ref{discuss}.

\section{Microscopic model}
\label{basemodel}

We consider $M$ flavors of conduction electrons, $c$, hopping on a lattice that are coupled locally and randomly to SYK islands on each lattice site (Fig.~\ref{Modelfig}a). The islands contain $N$ flavors of valence electrons, $f$, which interact among themselves in such a way that they realize SYK models. The Hamiltonian for our system is given by 
\begin{align}
&H = -t\sum_{\langle rr^\prime\rangle;~i=1}^M (c^\dg_{ri} c_{r^\prime i} + \mathrm{h.c.}) - \mu_c \sum_{r;~i=1}^M c^\dg_{ri} c_{ri}  - \mu \sum_{r;~i=1}^N f^\dg_{ri} f_{ri} \nn
&+ \frac{1}{NM^{1/2}}\sum_{r;~i,j=1}^N \sum_{k,l=1}^M g^r_{ijkl}  f^\dg_{ri}f_{rj}c^\dg_{rk} c_{rl} + \frac{1}{N^{3/2}}\sum_{r;~i,j,k,l=1}^NJ^r_{ijkl} f^\dg_{ri}f^\dg_{rj}f_{rk}f_{rl}.
\label{ham}
\end{align}
We will take the limits of $M=\infty$ and $N=\infty$, but we will be interested in values of $M/N$ that are at most $\mathcal{O}(1)$. We choose $J^r_{ijkl}$ and $g^r_{ijkl}$ as independent complex Gaussian random variables, with $\ll J^r_{ijkl} J^{r^\prime}_{lkij}\gg = (J^2/8)\delta_{rr^\prime}$ and $\ll g^r_{ijkl} g^{r^\prime}_{jilk} \gg = g^2 \delta_{rr^\prime}$ and all other $\ll..\gg$'s being zero, where $\ll..\gg$ denotes disorder-averaging.
\change{Note that $t$ is non-random, and this will lead to a Fermi surface for the $c$ fermions.}
 The disorder-averaged action then is
\begin{align}
&S = \int_0^\beta d\tau \left[ \sum_{r;~i=1}^M c^\dg_{ri}(\tau)(\partial_\tau -\mu_c)c_{ri}(\tau)-t\sum_{\langle rr^\prime \rangle;~i=1}^M (c^\dg_{ri}(\tau) c_{r^\prime i}(\tau) + \mathrm{h.c.})+\sum_{r;~i=1}^N f^\dg_{ri}(\tau)(\partial_\tau-\mu)f_{ri}(\tau^\prime)\right] \nn
&-M\frac{g^2}{2}\sum_r\int_0^\beta d\tau d\tau^\prime G^c_r(\tau-\tau^\prime)G^c_r(\tau^\prime-\tau) G_r(\tau-\tau^\prime)G_r(\tau^\prime-\tau) \nn
&-N\frac{J^2}{4}\sum_r \int_0^\beta d\tau d\tau^\prime G_r^2(\tau-\tau^\prime)G_r^2(\tau^\prime-\tau) -N\sum_r \int_0^\beta d\tau d\tau^\prime \Sigma_r(\tau-\tau^\prime)\left(G_r(\tau^\prime-\tau)+\frac{1}{N}\sum_{i=1}^Nf^\dg_{ri}(\tau)f_{ri}(\tau^\prime)\right) \nn
&-M\sum_r\int_0^\beta d\tau d\tau^\prime \Sigma^c_r(\tau-\tau^\prime)\left(G^c_r(\tau^\prime-\tau)+\frac{1}{M}\sum_{i=1}^Mc^\dg_{ri}(\tau)c_{ri}(\tau^\prime)\right),
\end{align}
where we have followed the usual strategy for SYK models~\cite{Sachdev2015,Sachdev2017} and introduced the auxiliary fields $G,\Sigma,G^c,\Sigma^c$ corresponding to Green's functions and self-energies of the $f$ and $c$ fermions respectively at each lattice site. In the $M,N=\infty$ limit, the integrals over the $\Sigma,\Sigma^c$ fields enforce the definitions of $G,G^c$ at each lattice site $r$. The large $M$, $N$ saddle-point equations are obtained by varying the action with respect to these $G$ and $\Sigma$ fields after integrating out the fermions
\begin{align}
&\Sigma_r(\tau-\tau^\prime) = \Sigma(\tau-\tau^\prime) = - J^2 G_r^2(\tau-\tau^\prime)G_r(\tau^\prime-\tau) - \frac{M}{N} g^2 G_r(\tau-\tau^\prime)G_r^c(\tau-\tau^\prime)G_r^c(\tau^\prime-\tau) \nn
&= - J^2 G^2(\tau-\tau^\prime)G(\tau^\prime-\tau) - \frac{M}{N} g^2 G(\tau-\tau^\prime)G^c(\tau-\tau^\prime)G^c(\tau^\prime-\tau), \nn
&G(i\omega_n) = \frac{1}{i\omega_n + \mu - \Sigma(i\omega_n)},
\label{Dysonsaddle0}
\end{align}
\change{and}
\begin{align}
&\Sigma^c_r(\tau-\tau^\prime) = \Sigma^c(\tau-\tau^\prime) = -g^2G^c_r(\tau-\tau^\prime)G_r(\tau-\tau^\prime)G_r(\tau^\prime-\tau) = -g^2G^c(\tau-\tau^\prime)G(\tau-\tau^\prime)G(\tau^\prime-\tau), \nn
&G^c(i\omega_n) = \int \frac{d^dk}{(2\pi)^d} \frac{1}{i\omega_n - \epsilon_k + \mu_c - \Sigma^c(i\omega_n)} \equiv \int \frac{d^dk}{(2\pi)^d}G^c(k,i\omega_n).
\label{Dysonsaddle}
\end{align}

\change{The last expression shows that the $c$ fermions have a dispersion $\epsilon_k$ and an associated Fermi 
surface; the lifetime of the Fermi surface excitations will be determined by the frequency dependence of $\Sigma^c$, which will be computed in the next section.} We define chemical potentials such that half-filling occurs when $\mu=\mu_c=0$. The islands are not capable of exchanging electrons with the Fermi sea, so there is no reason {\it a priori} to have $\mu=\mu_c$, or even for islands at different sites to have the same $\mu$. However, for convenience we will keep the $\mu$ of all the islands the same. \change{The} real system \change{would} operate at fixed {\it densities}, and $\mu$ and $\mu_c$ will appropriately renormalize as the mutual coupling $g$ is varied, in order to keep the densities of $c$ and $f$ individually fixed, \change{as the interaction between $c$ and $f$ conserves their numbers individually}. However, as we shall find, the half-filled case always corresponds to $\mu=\mu_c=0$ regardless of $g$. We will always have $J\gg T$ in this work, and also $J\gtrsim g$. A sketch of the phases realized by our model as a function of temperature is shown in Fig.~\ref{Modelfig}b.

\section{Fate of the conduction electrons}
\label{mflandim}

\subsection{The case of infinite bandwidth}
\label{infiniband}

We first consider the case of infinite bandwidth, or equivalently $t \gg g, J \gg T$. The precise value of $\mu_c$ doesn't matter as long as its magnitude is not infinite, as the conduction electrons float on an effectively infinitely deep Fermi sea. Then, we can use the standard trick for evaluating integrals about a Fermi surface, and we have
\beq
G^c(i\omega_n) = \int \frac{d^dk}{(2\pi)^d} \frac{1}{i\omega_n - \epsilon_k + \mu_c - \Sigma^c(i\omega_n)} \rightarrow \nu(0)\int_{-\infty}^\infty\frac{d\varepsilon}{2\pi}\frac{1}{i\omega_n - \varepsilon - \Sigma^c(i\omega_n)},
\eeq
where $\nu(0)$ is the density of states at the Fermi energy. 

We take the lattice constant $a$ to be $1$. This makes $k$ dimensionless by redefining $ka$ to be $k$. The energy dimension of $\epsilon_k$ then comes from the inverse band mass. The density of states $\nu(0)$ then has the dimension of 1/(energy) (on a lattice $\nu(0)\sim1/t\sim1/\Lambda$, where $\Lambda$ is the bandwidth).

We will also have $\mathrm{sgn}(\mathrm{Im}[\Sigma^c(i\omega_n)])  = -\mathrm{sgn}(\omega_n)$, so
\beq
G^c(i\omega_n) = -\frac{i}{2}\nu(0)\mathrm{sgn}(\omega_n),~~ G^c(\tau) =  -\frac{\nu(0)T}{2\sin(\pi T\tau)},~~-\beta \le \tau \le \beta,
\eeq
with other intervals obtained by applying the Kubo-Martin-Schwinger (KMS) condition $G^c(\tau+\beta)=-G^c(\tau)$. At $T=0$, we have 
\beq
G^c(\tau,T=0) = -\frac{\nu(0)}{2\pi\tau}.
\label{gc0}
\eeq

We consider $M/N=0$ to begin with. Then, the $f$ electrons are not affected by the $c$ electrons, and their Green's functions are exactly 
\change{of the incoherent form} of the SYK model, which, in the low-energy limit, are given by~\cite{Sachdev2015} 
\beq
G(\tau) = -\frac{\pi^{1/4}\cosh^{1/4}(2\pi\mathcal{E})}{J^{1/2}\sqrt{1+e^{-4\pi\mathcal{E}}}}\left(\frac{T}{\sin(\pi T \tau)}\right)^{1/2}e^{-2\pi\mathcal{E}T\tau},~~0\le \tau < \beta
\label{sykconf}
\eeq
where $\mathcal{E}$ is a function of $\mu$ with $\mathcal{E} \propto -\mu/J$ for small $\mu/J$. Other intervals are again obtained by the KMS condition $G(\tau+\beta)=-G(\tau)$. The zero-temperature limit of this, and similar expressions appearing later, can be straightforwardly taken~\cite{Sachdev2015}
\beq
G(\tau>0,T=0) = -\frac{\cosh^{1/4}(2\pi\mathcal{E})}{\pi^{1/4}J^{1/2}\sqrt{1+e^{-4\pi\mathcal{E}}}}\frac{1}{\tau^{1/2}},~~G(\tau<0,T=0) = \frac{\cosh^{1/4}(2\pi\mathcal{E})}{\pi^{1/4}J^{1/2}\sqrt{1+e^{4\pi\mathcal{E}}}}\frac{1}{|\tau|^{1/2}}
\label{sykconf0}
\eeq

\change{
Now we can compute the self energy of the $c$ fermions, which is}
\beq
\Sigma^c(\tau) = -g^2G^c(\tau)G(\tau)G(-\tau) = -\frac{\pi^{1/2}g^2\nu(0)T^2}{4J\cosh^{1/2}(2\pi\mathcal{E})\sin^2(\pi T \tau)},~~0\le \tau < \beta.
\eeq
Fourier transforming with a cutoff of $\tau$ at $J^{-1}\ll T^{-1}$ and $\beta-J^{-1}$ gives
\beq
\Sigma^c(i\omega_n) = \frac{ig^2\nu(0)T}{2J\cosh^{1/2}(2\pi\mathcal{E})\pi^{3/2}}\left(\frac{\omega_n}{T} \ln \left(\frac{2\pi  Te^{\gamma_E -1}}{J}\right)+\frac{\omega_n}{T} \psi\left(\frac{\omega_n}{2 \pi T}\right)+\pi \right),
\label{sigf}
\eeq
where $\psi$ is the digamma function and $\gamma_E$ is the Euler-Mascheroni constant. As foreseen, this satisfies $\mathrm{sgn}(\mathrm{Im}[\Sigma^c(i\omega_n)])  = -\mathrm{sgn}(\omega_n)$ on the fermionic Matsubara frequencies. For $|\omega_n|\gg T$
\beq
\Sigma^c(i\omega_n) \rightarrow \frac{ig^2\nu(0)}{2J\cosh^{1/2}(2\pi\mathcal{E})\pi^{3/2}}\omega_n\ln\left(\frac{|\omega_n|e^{\gamma_E-1}}{J}\right).
\label{sigf0}
\eeq
\change{Note the MFL form of the itinerant $c$ fermion self energy, $\sim \omega \ln \omega$}.
Since the large $N$ and $M$ limits are taken at the outset, this \change{MFL} is stable even as $T\rightarrow0$. For finite $N$ and $M$, the coupling $g$ is irrelevant in the infrared (IR)~\cite{McGreevy2017}, and the model reduces to a theory of non-interacting electrons as $T\rightarrow0$, with the MFL existing only above a temperature scale whose magnitude is suppressed in $N$ and the zero-temperature entropy going to zero.

Upon analytically continuing $i\omega_n\rightarrow \omega+i0^+$, we get the inverse lifetime for the conduction electrons defined by
\beq
\gamma \equiv  -2\mathrm{Im}[\Sigma^c_R(0)] \equiv -\mathrm{Im}[\Sigma^c(i\omega_n\rightarrow 0+i0^+)] = \frac{g^2\nu(0)T}{J\cosh^{1/2}(2\pi\mathcal{E})\pi^{1/2}}.
\label{dumprate}
\eeq
Since the coupling of the conduction electrons to the SYK islands is spatially disordered, this rate also represents the transport scattering rate up to a constant numerical factor. The scattering of $c$ electrons off the islands requires the $f$ electrons inside the islands to move between orbitals. Hence $\gamma$ vanishes when the islands are flooded or drained by sending $\mathcal{E}\rightarrow\mp \infty$ respectively, say, by doping them.  

If we do not have $M/N=0$, the SYK Green's function will be affected as there is a back-reaction self-energy to the SYK islands. To see what this does when we perturbatively turn on $M/N$, we compute it with the $M/N=0$ Green's functions with a cutoff of $\tau$ at $J^{-1}$ and $\beta-J^{-1}$
\beq
\tilde{\Sigma}(\tau) = -\frac{M}{N}g^2G(\tau)G^c(\tau)G^c(-\tau) \approx -\frac{M\pi^{1/4}\cosh^{1/4}(2\pi\mathcal{E})g^2\nu^2(0)T^{5/2}e^{-2\pi\mathcal{E}T\tau}}{4NJ^{1/2}\sqrt{1+e^{-4\pi\mathcal{E}}}\sin^{5/2}(\pi T \tau)}.
\eeq
If $\mathcal{E}=0$, then $\tilde{\Sigma}(i\omega_n) \propto i(M/N)g^2\nu^2(0)\omega_n$ as $T,\omega_n\rightarrow 0$, which is sub-leading to $\Sigma(i\omega_n)|_{M/N=0}\sim (J\omega_n)^{1/2}$, so the SYK character of the islands survives in the IR. 

\change{Now we consider the case of particle-hole symmetry breaking with a non-zero spectral asymmetry, $\mathcal{E}$ in Eq.~(\ref{Glocal}); we will find
that the basic structure of the results described above persists.}
If $\mathcal{E}\neq0$ but is small, then for $T\rightarrow0$, $\tilde{\Sigma}(i\omega_n\rightarrow0) \sim  -(M/N)g^2\nu^2(0)J\mathcal{E} \propto (M/N)g^2\nu^2(0)\mu+\mathcal{O}(i\omega_n)$. In contrast $\Sigma(i\omega_n\rightarrow0)|_{M/N=0} \sim \mu + \mathcal{O}(\omega_n^{1/2})$. Therefore the frequency-dependent part of $\tilde{\Sigma}$ is still subleading. Hence, in the IR we may still assume that all that happens to the SYK islands is that their chemical potential $\mu$ gets renormalized. By solving $\mathrm{Re}[\Sigma(i\omega_n\rightarrow0,T=0)]=\mu$, we obtain the corrected $\mathcal{E}\leftrightarrow\mu$ relation. At small $\mu/J$, this is
\beq
\mathcal{E} \approx -\frac{\mu/J}{\pi^{1/4}\sqrt{2}\left(1+ \displaystyle \frac{g^2\nu^2(0)M}{6\pi^{3/2}N}\right)}.
\label{newemu}
\eeq
The total particle number on each island, $\mathcal{N}_r = \sum_i f^\dg_{ir}f_{ir}$, commutes with $H$. Since the SYK particle density $\mathcal{Q}=\mathcal{N}/N$ is a universal function of $\mathcal{E}$, independent of $\mu$ and $J$, (\ref{newemu}) just implies a renormalization of the nonuniversal UV parts of the SYK Green's function and the island chemical potential, while the particle density remains fixed. Similarly, the vanishing of the zero-frequency real part of (\ref{sigf}) regardless of $\mathcal{E}$ implies that there is no renormalization of either the density or chemical potential of the conduction electrons in this infinite-bandwidth limit, since their number is independently conserved as well. For a finite bandwidth, the chemical potential of the conduction electrons renormalizes in such a way that their density remains fixed. 

In Appendix~\ref{pairhop}, we consider the effects of adding a `pair-hopping' term to~(\ref{ham}),
\beq
H \rightarrow H + \frac{1}{NM^{1/2}}\sum_{r;~i,j=1}^N\sum_{k,l=1}^M\left[\eta^r_{ijkl}f^\dg_{ri}f^\dg_{rj}c_{rk}c_{rl}+\mathrm{h.c.}\right],
\label{hph}
\eeq
with $\ll |\eta^r_{ijkl}|^2 \gg = \eta^2/8$, and $J\gtrsim\eta$. This term has identical power-counting to the $f^\dg f c^\dg c$ term, but can trade $c$ electrons for $f$ electrons and vice-versa. Since the numbers of $c$ and $f$ electrons are no longer independently conserved in this case, there is only one chemical potential, and $\mu_c=\mu$. We find that this term also generates an MFL as long as the bandwidth of the $c$ electrons is large.

As is well known, the marginal-Fermi liquid self-energy we obtained (\ref{sigf},~\ref{sigf0}) also leads to the leading low-temperature contribution to the specific heat \change{coming from the itinerant electrons} scaling as $C_V^{\mathrm{MFL}} \sim M g^2 (\nu(0))^2 (T/J)\ln (J/T)$~\cite{Crisan1996}. Note that the entropy has a non-vanishing $T \rightarrow 0 $ limit from the contribution of the SYK islands in the limit of $N\rightarrow\infty$ \cite{GPS01}, but this does not contribute to the specific heat. \change{The contribution to the specific heat coming from the SYK islands scales linearly in $T$ as $T\rightarrow0$~\cite{Sachdev2017}, which is subleading to the $T\ln T$ contribution of the itinerant electrons}.        

\subsection{The case of a finite bandwidth}
\label{dialup}

\change{
This subsection will show that a finite bandwidth does not modify the basic structure of the low-temperature MFL phase described above. However, if interactions \change{between $c$ and $f$} are strong enough, a crossover into an IM phase is possible at higher temperatures. Readers not interested in the details of the arguments can move ahead to the next section.}

If the bandwidth (and hence Fermi energy) of the conduction electrons is sizeable compared to the couplings, then the \change{momentum-integrated} local Green's function $G^c(i\omega_n)$ is no longer independent of the details of the self energy $\Sigma^c(i\omega_n)$. We consider two spatial dimensions, with the isotropic dispersion $\varepsilon_k = k^2/(2m)-\Lambda/2$, and a bandwidth $\varepsilon_k^{\mathrm{max}}-\varepsilon_{k=0} = \Lambda$.  Since $k$ is dimensionless, the band mass $m$ has dimensions of $1/(\mathrm{energy})$. The density of states is then just $\nu(\varepsilon)=\nu(0)=m$, at all energies $\varepsilon$, and we implicitly make use of this fact while simplifying and rewriting certain expressions. On a lattice, $m\sim \nu(0) \sim 1/t\sim 1/\Lambda$. 

The momentum-integrated conduction electron Green's function is
\beq
G^c(i\omega_n) = \frac{\nu(0)}{2\pi}\left[\ln(\Lambda +2\mu_c+2i\omega_n-2\Sigma^c(i\omega_n))-\ln(2\mu_c-\Lambda+2i\omega_n-2\Sigma^c(i\omega_n))\right].
\label{fbwarctan}
\eeq
We still expect $\mathrm{sgn}(\mathrm{Im}[\Sigma^c(i\omega_n)])=-\mathrm{sgn}(\omega_n)$. The chemical potential $\mu_c$ must now take an appropriate value to reproduce the correct density of conduction electrons. The conduction band filling is given by 
\beq
\mathcal{Q}_c = \frac{2\pi G^c(\tau=0^-)}{\nu(0)\Lambda}, 
\eeq
for the exact solution to $G^c$, which can be found by the imaginary-time \verb|MATLAB| code \verb|ggc.m|~\cite{Code} (The low-energy `conformal-limit' solutions described below are not valid at the short times $0^-$, and do not display this property).  

In general, the Dyson equations can now only be solved numerically, which the imaginary-time \verb|MATLAB| code \verb|ggc.m|~\cite{Code} and real-time \verb|MATLAB| code \verb|ggcrealtime.m|~\cite{CodeRT} do, albeit by holding the chemical potentials $\mu$ and $\mu_c$, rather than densities, fixed. In an extreme limit where $|i\omega_n+\mu_c-\Sigma^c(i\omega_n)|$ far exceeds the bandwidth for all $\omega_n$, which can happen only at $T\neq0$, we have a simplification of (\ref{fbwarctan}), obtained by expanding in $\Lambda$,
\beq
G^c(i\omega_n) = \frac{\Lambda \nu(0)}{2\pi(i\omega_n+\mu_c-\Sigma^c(i\omega_n))}.
\label{sfinc}
\eeq
This then leads to an SYK solution in the low-energy conformal limit for both $G$ and $G^c$, realizing a fully incoherent metal. We use the trial solutions
\beq
G^c(\tau) = - \frac{C_c}{\sqrt{1+e^{-4\pi\mathcal{E}_c}}}\left(\frac{T}{\sin(\pi T\tau)}\right)^{1/2}e^{-2\pi\mathcal{E}_cT\tau},~~G(\tau) = - \frac{C}{\sqrt{1+e^{-4\pi\mathcal{E}}}}\left(\frac{T}{\sin(\pi T\tau)}\right)^{1/2}e^{-2\pi\mathcal{E}T\tau},~~0\le \tau < \beta.
\label{gfincconf}
\eeq 
$\mathcal{E}_c$ is universally related to the conduction band filling, with $\mathcal{E}_c=0$ at half filling, and $\mathcal{E}_c\rightarrow\mp\infty$ when the band is full or empty respectively. When $M/N = 0$, there is no back-reaction to the islands, and $G$ is given by (\ref{sykconf}). We use the conditions $\mathrm{Re}[\Sigma^c(i\omega_n\rightarrow0,T=0)]=\mu_c$ and $G^c(i\omega_n\rightarrow0,T=0)=\Lambda\nu(0)/(2\pi(\mu_c-\Sigma^c(i\omega_n\rightarrow0,T=0)))$ to determine $C_c$, and also $\mu_c$ in terms of the fixed $\mathcal{E}_c$. Cutting off $\tau$ integrals in the Fourier transforms at a distance $\alpha_{\mathrm{UV}}^{-1}$ from singularities, we have 
\beq
C_c  = \frac{\cosh^{1/4}(2\pi\mathcal{E})}{2^{1/2}\pi^{1/4}J_{\mathrm{IM}}^{1/2}},~~J_{\mathrm{IM}}\equiv \frac{g^2}{J\Lambda\nu(0)}~~\mathrm{and}~~\mathcal{E}_c \approx -\frac{\pi^{1/4}\cosh^{1/4}(2\pi\mathcal{E})\mu_c}{g\Lambda^{1/2}\nu^{1/2}(0)}\sqrt{\frac{J}{\alpha_{\mathrm{UV}}}}~~(\mathrm{At~small}~\mu_c/g),
\eeq
with no feedback on the SYK islands. For~(\ref{sfinc}) to derive from~(\ref{fbwarctan}), this requires $|\mu_c-\Sigma^c(i \omega_n\rightarrow0)|\gg \Lambda$ or 
\beq
T\gg T_{\mathrm{inc}} \equiv \frac{\Lambda J}{\nu(0)g^2}.
\label{tinc}
\eeq
Furthermore, for (\ref{sykconf}) and (\ref{gfincconf}) to hold, we also need $J\gg T_{\mathrm{inc}}$ and $J_{\mathrm{IM}}\gg T_{\mathrm{inc}}$, implying $g^2\gg \Lambda J$. For $T\ll T_{\mathrm{inc}}$, we go back to the MFL, which now has a UV cutoff of $T_{\mathrm{inc}}$ instead of $J$, with its self energy going as $\Sigma^c(i\omega_n)\sim (g^2\nu(0)/J)i\omega_n\ln(|\omega_n|/T_{\mathrm{inc}})$. The choice of the UV cutoff $\alpha_{\mathrm{UV}}$ in the IM only affects the nonuniversal $\mathcal{E}_c\leftrightarrow\mu_c$ relation. An appropriate choice of the cutoff is $\alpha_{\mathrm{UV}}\sim J_{\mathrm{IM}} \lesssim J$. 

Turning on a small but finite $M/N$, we have to additionally use the conditions $\mathrm{Re}[\Sigma(i\omega_n\rightarrow0,T=0)]=\mu$ and $G(i\omega_n\rightarrow0,T=0)=1/(\mu-\Sigma(i\omega_n\rightarrow0,T=0))$ simultaneously to determine a renormalized $C$ and renormalized $\mu$, while keeping $\mathcal{E}$ fixed as before. We again cut off $\tau$ integrals in the Fourier transforms at a distance $\alpha_{\mathrm{UV}}^{-1}$ from singularities. This gives
\beq
C = \cosh^{1/4}(2\pi\mathcal{E})\frac{\pi^{1/4}}{J^{1/2}}\left(1-\frac{M}{N}\frac{\Lambda\nu(0)}{2\pi}\frac{\cosh(2\pi\mathcal{E})}{\cosh(2\pi\mathcal{E}_c)}\right)^{1/4},~~C_c = \frac{\cosh^{1/2}(2\pi\mathcal{E})\Lambda^{1/2}\nu^{1/2}(0)}{2^{1/2}Cg},
\label{ccfmn}
\eeq
and we do not show the nonuniversal $\mathcal{E},\mathcal{E}_c\leftrightarrow\mu,\mu_c$ relations because they are rather uninsightful and the physics is better described in terms of $\mathcal{E},\mathcal{E}_c$ which universally represent the conserved densities. 

If $M/N$ is increased to approach $(2\pi\cosh(2\pi\mathcal{E}_c))/(\Lambda \nu(0)\cosh(2\pi\mathcal{E}))$, the condition for incoherence that $|i\omega_n+\mu_c-\Sigma^c(i\omega_n)|$ exceed the bandwidth \change{for all $\omega_n$} becomes harder to fulfill, and larger and larger values of the coupling $g$ are required to achieve the IM phase at high temperatures. 

When $M/N> (2\pi\cosh(2\pi\mathcal{E}_c))/(\Lambda \nu(0)\cosh(2\pi\mathcal{E}))$, we still recover the MFL deep enough in the IR, due to the back-reaction self energy $\tilde{\Sigma}$ being irrelevant, and the conduction electron self energy $\Sigma^c$ also vanishing at the lowest energies. However, at values of the coupling $g$ large enough so that effects of the conduction electron bandwidth may be ignored above a certain temperature, we find a crossover into a different IM phase, with local Green's functions given by (at half-filling)
\beq
G^c(\tau) \sim \left(\frac{T}{\sin(\pi T\tau)}\right)^{\Delta_c},~~G(\tau) \sim \left(\frac{T}{\sin(\pi T\tau)}\right)^{1-\Delta_c},~~0<\Delta_c<1/2,
\label{GIM2}
\eeq
with $\Delta_c$ given by the solution to the equation 
\beq
\left(\frac{\Delta_c}{1-\Delta_c}\right)\cot^2\left(\frac{\pi\Delta_c}{2}\right)=\frac{M}{N}\frac{\Lambda\nu(0)}{2\pi},
\eeq
which has the property that $\Delta_c\rightarrow0$ as $M/N\rightarrow\infty$ and $\Delta_c\rightarrow1/2$ as $M/N\rightarrow 2\pi/(\Lambda \nu(0))$. These Green's functions may be derived by solving the Dyson equations (\ref{Dysonsaddle0}, \ref{Dysonsaddle}) while ignoring both the conduction electron dispersion and the coupling $J$. Indeed, with the scalings in (\ref{GIM2}), the term proportional to $J^2$ in the expression for $\Sigma(\tau)$ is irrelevant compared to the other term. This phase has a resistivity that scales as $T^{2(1-\Delta_c)}$. Since we are only interested in models with linear-in-$T$ resistivities, we will henceforth assume that $M/N$ is small enough to avoid this regime.

Since  $\nu(0)\sim1/\Lambda\sim 1/t$ on a lattice, fine-tuning $g\sim J\sim \Lambda \gg T$ makes the scattering rate~(\ref{dumprate}) `Planckian', i.e. an $\mathcal{O}(1)$ number times $T$, since it is given by {\it ratios} of large quantities. The MFL doesn't break down if we do this; In~(\ref{fbwarctan}), $|\Sigma^c(i(\omega_n\sim T))|\sim T\ln T/J \ll \Lambda$, so the infinite-bandwidth result~(\ref{dumprate}) is still applicable. The crossover to the IM doesn't occur either, since $T\ll T_\mathrm{inc}$, and finally, the part of the back-reaction self-energy to the SYK islands that does not renormalize their chemical potentials is $|\tilde{\Sigma}(i(\omega_n\sim T))]| \sim (M/N)(g\nu(0))^2 T$ which is $\ll |\Sigma(i(\omega_n\sim T))| \sim (JT)^{1/2}$, i.e. the part of the internal self-energy of the SYK islands that doesn't renormalize chemical potential, as long as $M/N$ is not $\gg 1$, so the SYK character of the islands also survives.

In the IM regime, since both the conduction and island electrons have local SYK Green's functions, the specific heat scales as $C_V^\mathrm{IM} \sim M T/J_{\mathrm{IM}} + N T/J$, with no logarithmic corrections~\cite{Sachdev2017}.

\section{Transport in a single domain}
\label{Transport}

\change{In this section we consider transport in two spatial dimensions, with the isotropic dispersion $\varepsilon_k = k^2/(2m)-\Lambda/2$. 
We will find that many aspects of the transport can be computed in a traditional Boltzmann transport computation, due to the large $N$ and $M$ limits. In particular, 
quantum corrections to transport, of the type leading to quantum interference and localization, are suppressed by the local disorder, the non-quasiparticle nature of the charge carriers, and the large number of fermion flavors.}

\begin{figure}
\begin{center}
\includegraphics[height=2.0in]{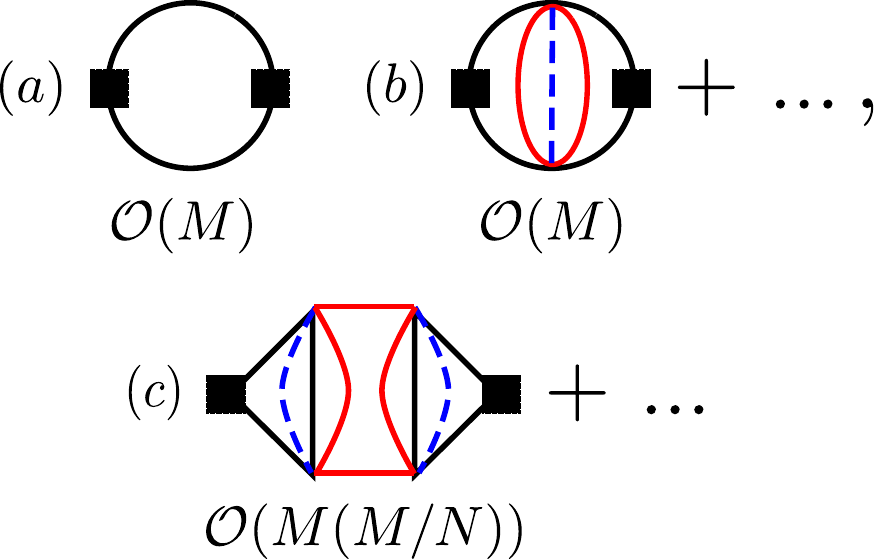}
\end{center}
\caption{(a) The uniform current-current correlation bubble used to compute conductivities. The current vertices are black squares and the black lines are conduction electron ($c$) propagators. (b) and (c) Additional diagrams forming ladder series, with ladder units of up to $3$ loops, that contribute to the conductivities and are not immediately suppressed by the large $N$ and $M$ limits. The red lines are island fermion ($f$) propagators that do not carry momentum. The dashed blue lines carry momentum and come from disorder averaging of the non-translationally invariant coupling $g^x_{ijkl}$. These diagrams however vanish upon momentum integration in the loops containing the current vertices, for reasons mentioned in the main text.}
\label{Goodbadugly}
\end{figure}

In our double large $N$ and $M$ limit, if $M/N=0$, the only vertex corrections to the uniform conductivities that aren't trivially killed by this limit are the ones that involve uncrossed vertical ladders of $f^\dg_if_j$ propagators in the current-current correlator bubbles (First diagram of Fig.~\ref{Goodbadugly}b). However, since the $f$ propagators are purely local and independent of momentum, these diagrams vanish due to averaging of the vector velocity in the current vertices over the closed fixed-energy contours in momentum space, as the scattering of the conduction electrons is isotropic, just like in the textbook problem of the non-interacting disordered metal~\cite{Efros2012}. Unlike the non-interacting disordered metal, there is no localization in two dimensions as the crossed-ladder `Cooperon' diagrams are suppressed by the large $M$ limit. Hence, the relaxation-time-like approximation of keeping only self-energy corrections is valid. 

If $M/N$ is nonzero but $\mathcal{O}(1)$ or smaller, then certain 3-loop and higher order ladder insertions (Such as Fig.~\ref{Goodbadugly}c) also contribute extensively in $M$ to the current-current correlation. However, these diagrams again vanish due to the averaging of the vector velocity mentioned above. All this happens regardless of the values of $g,J,\Lambda,\mu_c$, and for both energy and electrical currents. 

\subsection{Marginal-Fermi liquid}
\label{mfltransport}

We first discuss \change{a Boltzmann transport approach in} the MFL regime. For simplicity, we consider infinite bandwidth and an infinitely deep Fermi sea. The uniform current-current correlation bubble (Fig.~\ref{Goodbadugly}a) is given by, for an isotropic Fermi surface,
\beq
\langle I_x I_x \rangle(i\Omega_m) = - M\frac{v_F^2}{2}\nu(0)T\sum_{\omega_n}\int_{-\infty}^\infty\frac{d\varepsilon}{2\pi}\frac{1}{i\omega_n-\varepsilon-\Sigma^c(i\omega_n)}\frac{1}{i\omega_n+i\Omega_m-\varepsilon-\Sigma^c(i\omega_n+i\Omega_m)},
\eeq
where $v_F = k_F/m$ is the Fermi velocity (on a lattice $v_F\sim t$, since the lattice constant $a$ is set to $1$). Using the spectral representation, this can be converted to give the DC conductivity
\beq
\sigma_0^{\mathrm{MFL}} = M\frac{v_F^2\nu(0)}{16T}\int_{-\infty}^\infty\frac{dE_1}{2\pi}\mathrm{sech}^2\left(\frac{E_1}{2T}\right)\frac{1}{|\mathrm{Im}\Sigma_R^c(E_1)|}.
\eeq
Inserting the self energy, we can scale out $T$ and numerically evaluate the integral, giving
\beq
\sigma_0^{\mathrm{MFL}} = 0.120251\times MT^{-1}J\times\left(\frac{v_F^2}{g^2}\right)\cosh^{1/2}(2\pi\mathcal{E}).
\label{s0MFL}
\eeq
If we want $\sigma_0^{\mathrm{MFL}}/M \ll 1$, we must have $T\gg T_{\mathrm{inc}}$, implying a crossover into the IM regime. Thus the MFL is never a true bad metal, but its resistivity can still numerically exceed the quantum unit $h/e^2$, depending on parameters. 

The `open-circuit' thermal conductivity $\kappa_0^{\mathrm{MFL}}$, which is defined under conditions where no electrical current flows,  is given by
\beq
\kappa_0^{\mathrm{MFL}} = \bar{\kappa}_0^{\mathrm{MFL}} - \frac{(\alpha_0^{\mathrm{MFL}})^2T}{\sigma_0^{\mathrm{MFL}}},
\eeq
where $\bar{\kappa}_0^{\mathrm{MFL}}$ is the `closed-circuit' thermal conductivity in the presence of electrical current, and $\alpha_0^{\mathrm{MFL}}$ is the thermoelectric conductivity. The thermoelectric conductivity vanishes when the temperature is much smaller than the bandwidth and Fermi energy, due to effective particle-hole symmetry about the Fermi surface, so $\kappa_0^{\mathrm{MFL}} = \bar{\kappa}_0^{\mathrm{MFL}}$. The Lorenz ratio is then given by
\beq
L ^{\mathrm{MFL}}= \frac{\kappa_0^{\mathrm{MFL}}}{\sigma_0^{\mathrm{MFL}}T} =\frac{\bar{\kappa}_0^{\mathrm{MFL}}}{\sigma_0^{\mathrm{MFL}}T} = \frac{\int_{-\infty}^\infty\frac{dE_1}{2\pi}E_1^2\mathrm{sech}^2\left(\frac{E_1}{2}\right)\frac{1}{|\mathrm{Im}[E_1\psi(-iE_1/(2\pi))+i\pi]|}}{\int_{-\infty}^\infty\frac{dE_1}{2\pi}\mathrm{sech}^2\left(\frac{E_1}{2}\right)\frac{1}{|\mathrm{Im}[E_1\psi(-iE_1/(2\pi))+i\pi]|}} =  0.713063\times L_0,
\eeq
which is smaller than $L_0=\pi^2/3$ for a Fermi liquid.

In the presence of a uniform transverse magnetic field, we can use the following improved relaxation-time linearized Boltzmann equation (which incorporates an off-shell distribution function) for a temporally slowly-varying and spatially uniform applied electric field~\cite{Kamenev2011,Nave2007}, since there are no Cooperons in the large-$M$ limit, and hence none of the typical localization-related corrections~\cite{Altshuler1980} to the conductivity tensor. The Boltzmann equation reads (here, $t$ is time, not the hopping amplitude, and $\mathcal{B}$ is a dimensionless version of the magnetic field $B$ which shall be explained below)
\beq
(1-\partial_\omega\mathrm{Re}[\Sigma^c_R(\omega)])\partial_t\delta n(t,k,\omega)  + v_F \hat{k}\cdot \mathbf{E}(t)~n_f^\prime(\omega) + v_F (\hat{k}\times \mathcal{B}\hat{z})\cdot\nabla_k \delta n(t,k,\omega)  = 2\delta n(t,k,\omega) \mathrm{Im}[\Sigma^c_R(\omega)],
\label{BE1}
\eeq
where $n_f(\omega)=1/\left(e^{\omega/T}+1\right)$ is the Fermi distribution, $\delta n$ is the change in the distribution due to the applied electric field, the conduction electrons are negatively charged, and the magnetic field points out of the plane of the system. This equation is derived in Appendix~\ref{BZE} from the Dyson equation on the Keldysh contour, and can be solved by the ansatz $\delta n(t,k,\omega) = k\cdot\varphi(t,\omega) = k_i \varphi_i(t,\omega)$. 

In the DC limit, the effective mass enhancement $(1-\partial_\omega\mathrm{Re}[\Sigma^R(\omega)])$ does not matter~\cite{Nave2007} (the effective mass enhancement is important for AC magnetotransport and affects the frequency at which the cyclotron resonance occurs; it shifts the cyclotron resonance from the cyclotron frequency defined by the bare mass to the one defined by the effective mass. The enhanced effective mass also appears in the specific heat~\cite{Crisan1996} and Lifshitz-Kosevich formula~\cite{Pelzer1991} of MFLs). We then have
\beq
v_F \hat{k}\cdot \mathbf{E}~n_f^\prime(\omega) +v_F (\hat{k}\times \mathcal{B}\hat{z})\cdot\nabla_k \delta n(k,\omega) = 2\delta n(k,\omega) \mathrm{Im}[\Sigma_R^c(\omega)],
\label{IBE}
\eeq

We note that in~(\ref{IBE}), $\mathcal{B}$ is dimensionless in our choice of units. Since the quantities we set to $1$ were the magnitude of the electron charge $e$, the lattice constant $a$, and $\hbar$ and $k_B$, we have 
\beq
\mathcal{B} = \frac{eBa^2}{\hbar},
\label{BcalB}
\eeq
i.e. the flux per unit cell in units of $\hbar/e$. 

Substituting $\delta n(k,\omega) = k_i \varphi_i(\omega)$ into (\ref{IBE}), we obtain
\beq
\varphi_i(\omega) = \frac{v_F}{k_F}n_f^\prime(\omega)\left(2\mathrm{Im}[\Sigma^c_R(\omega)]\delta_{ij}+\epsilon_{ij}\mathcal{B}\frac{v_F}{k_F}\right)^{-1}_{ij}E_j.
\eeq
Using the current density
\beq
I_i = -M\nu(0)\int_0^{2\pi}\frac{d\theta}{2\pi}\int_{-\infty}^{\infty}\frac{d\omega}{2\pi}v_F\hat{k}_i \delta n(k_F\hat{k},\omega),
\eeq
we get the longitudinal and Hall conductivities
\begin{align}
&\sigma_L^{\mathrm{MFL}} = M\frac{v_F^2\nu(0)}{16T}\int_{-\infty}^{\infty}\frac{dE_1}{2\pi} \mathrm{sech}^2\left(\frac{E_1}{2T}\right)\frac{-\mathrm{Im}[\Sigma^c_R(E_1)]}{\mathrm{Im}[\Sigma^c_R(E_1)]^2+(v_F/(2k_F))^2\mathcal{B}^2}, \nonumber \\
&\sigma_H^{\mathrm{MFL}} = -M\frac{v_F^2\nu(0)}{16T}\int_{-\infty}^{\infty}\frac{dE_1}{2\pi} \mathrm{sech}^2\left(\frac{E_1}{2T}\right)\frac{(v_F/(2k_F))\mathcal{B}}{\mathrm{Im}[\Sigma^c_R(E_1)]^2+(v_F/(2k_F))^2\mathcal{B}^2}.
\label{mfllcmag}
\end{align}

Note that, \change{given the scaling of (\ref{sigf})}, these can be immediately written as 
\beq
\sigma_L^{\mathrm{MFL}}\sim T^{-1}s_L((v_F/k_F)(\mathcal{B}/T)),~~\sigma_H^{\mathrm{MFL}}\sim -\mathcal{B}T^{-2}s_H((v_F/k_F)(\mathcal{B}/T)). \label{eq:Bscale}
\eeq 
The asymptotic forms of the functions $s_L$ and $s_H$ are
\beq
s_{L,H}(x\rightarrow\infty) \propto 1/x^2,~~s_{L,H}(x\rightarrow 0) \propto x^0. 
\eeq
\change{So we have obtained the advertised $B/T$ scaling in the MFL regime. However, with the asymptotic forms noted above, it is not difficult
to see that the magnetoresistance, $\rho_{xx}$ saturates at large $B$. Nevertheless, the results above will be useful as inputs into our consideration
of the effects of macroscopic disorder in Section~\ref{EMARRN}: we will show there that the $B/T$ scaling survives, and the macroscopic disorder
leads to a linear in $B$ magnetoresistance.}

\change{We now show that the numerical scale of the $B/T$ crossover is in general accord with the observations.}
In (\ref{mfllcmag}), for the `Planckian' choice of parameters described at the end of Sec.~\ref{dialup}, $B$ becomes `large' (i.e., the cyclotron term in the denominators overwhelms $\mathrm{Im}[\Sigma^c_R(E_1)]$ for $|E_1|\lesssim T$, causing $\sigma_H^{\mathrm{MFL}}$ to start decreasing with increasing $B$), when $eBa^2/\hbar\gtrsim k_BT/t$. Using reasonable values of the lattice constant $a = 3.82~\mathrm{\AA}$ and the hopping $t = 0.25~\mathrm{eV}$,  the above inequality can also roughly be written as $\mu_B B \gtrsim k_B T$, where $\mu_B$ is the Bohr magneton, since $a^2e t/\hbar \approx 0.96 \mu_B$ for these parameters. 

In the  analysis of the IM regime to follow, there is no such notion of `large' magnetic fields; regardless of the value of $B$, the field-dependent corrections to the conductivity tensor remain much smaller than its zero-field value.

\subsection{Incoherent metal}
\label{incmettransport}

\change{
This subsection considers transport in the IM phase discussed earlier, in which the Fermi surface is washed out,  and shows quantitatively that the orbital effects of a magnetic field on charge transport are strongly suppressed irrespective of the strength of the field. The physical reason for this effect is that the effective mean-free-path of the electrons in the IM is less than a lattice spacing, with conduction occurring locally and incoherently across individual lattice bonds. The effect of the Lorentz force on the electrons is thus negligible. If the reader is uninterested in the details of the following computations, they may move on to the next section.}

In the IM regime we have
\beq
\sigma_0^{\mathrm{IM}} = \frac{M\Lambda^2}{32\pi T}\int_{-\infty}^{\infty}\frac{dE_1}{2\pi}\mathrm{sech}^2\left(\frac{E_1}{2T}\right)(A^c(k,E_1))^2.
\eeq
The spectral function is independent of $k$ in the IM, and we decoupled the momentum integral implicit in the above equation, generating a prefactor of $\Lambda\nu(0)/(2\pi)$. For simplicity we consider $M/N=0$ in this subsection. A small finite $M/N$ only rescales $G^c$, as shown by (\ref{ccfmn}, \ref{gfincconf}), and hence leads to no qualitative difference in any of the following results. We have 
\begin{align}
&A^c(k,E_1) \equiv \frac{2\pi}{\Lambda\nu(0)}A^c(E_1) \equiv -\frac{4\pi}{\Lambda\nu(0)}\mathrm{Im}[G^c(i\omega_n\rightarrow E_1+i0^+)]  \nn
&=-2\mathrm{Im}\Bigg[\frac{i(-1)^{3/4}\pi^{1/4}(i+e^{2\pi\mathcal{E}_c})J^{1/2}\cosh^{1/4}(2\pi\mathcal{E})}{gT^{1/2}\Lambda^{1/2}\nu^{1/2}(0)\sqrt{1+e^{4\pi\mathcal{E}_c}}} 
\frac{\Gamma \left(\frac{1}{4}-\frac{i(E_1-2\pi\mathcal{E}_cT)}{2 \pi  T}\right)}{\Gamma \left(\frac{3}{4}-\frac{i(E_1-2\pi\mathcal{E}_cT)}{2 \pi  T}\right)}\Bigg],
\end{align}
and we get
\beq
\sigma_0^{\mathrm{IM}} = (\pi^{1/2}/8)\times MT^{-1}J\times\left(\frac{\Lambda}{\nu(0) g^2}\right)\frac{\cosh^{1/2}(2\pi \mathcal{E})}{\cosh(2\pi\mathcal{E}_c)}.
\label{s0IM}
\eeq
Due to the IM existing only at temperatures above $T_{\mathrm{inc}}$, given by (\ref{tinc}), we always have $\sigma_0^{\mathrm{IM}}/M \ll 1$, which makes the IM a bad metal. \change{Note that the slope of the resistivity $\rho_0(T)=1/\sigma_0(T)$ vs temperature in the IM generically differs from that in the MFL by an $\mathcal{O}(1)$ number, as can be seen by comparing (\ref{s0MFL}) and (\ref{s0IM}).}

The Lorenz ratio in the IM is (here, the thermoelectric conductivity $\alpha_0^\mathrm{IM}$ does not vanish, so $\kappa_0^{\mathrm{IM}}$ and $\bar{\kappa}_0^{\mathrm{IM}}$ are distinct quantities)
\beq
L ^{\mathrm{IM}}= \frac{\int_{-\infty}^\infty\frac{dE_1}{2\pi}E_1^2\mathrm{sech}^2\left(\frac{E_1}{2}\right)(A^c(E_1))^2-\frac{[\int_{-\infty}^\infty\frac{dE_1}{2\pi}E_1\mathrm{sech}^2\left(\frac{E_1}{2}\right)(A^c(E_1))^2]^2}{\int_{-\infty}^\infty\frac{dE_1}{2\pi}\mathrm{sech}^2\left(\frac{E_1}{2}\right)(A^c(E_1))^2}}{\int_{-\infty}^\infty\frac{dE_1}{2\pi}\mathrm{sech}^2\left(\frac{E_1}{2}\right)(A^c(E_1))^2} =  \frac{3}{8}\times L_0,~~\mathrm{regardless~of}~\mathcal{E},\mathcal{E}_c.
\label{lorenzIM}
\eeq
This result was also obtained by a different method for the IM of Ref.~\onlinecite{Balents2017}, although they only analyzed the particle-hole symmetric case equivalent to $\mathcal{E}_c=0$.

Another dimensionless ratio that is interesting is the thermopower, i.e. the ratio of the thermoelectric to electrical conductivities, 
\beq
\mathcal{S}_0^{\mathrm{IM}}=\frac{\alpha_0^{\mathrm{IM}}}{\sigma_0^{\mathrm{IM}}} = \frac{\int_{-\infty}^\infty\frac{dE_1}{2\pi}E_1\mathrm{sech}^2\left(\frac{E_1}{2}\right)(A^c(E_1))^2}{\int_{-\infty}^\infty\frac{dE_1}{2\pi}\mathrm{sech}^2\left(\frac{E_1}{2}\right)(A^c(E_1))^2} = 2\pi\mathcal{E}_c.
\label{dsdqIM}
\eeq
This relationship between the thermopower and the spectral asymmetry $\mathcal{E}_c$ was also found in a different \change{model of coupled SYK islands} realized in Ref.~\onlinecite{Sachdev2017}. The ratios (\ref{lorenzIM}), (\ref{dsdqIM}) hold even for a finite small $M/N$, as the effect of a finite small $M/N$ is simply a rescaling of the Green's function $G^c$. 

Let us describe the fate of magnetotransport in the IM regime. On a lattice, we have $\Lambda\nu(0)\sim1$. Then $J_{\mathrm{IM}} = g^2/J$, and the conduction electron self-energy is $\sim \sqrt{J_{\mathrm{IM}}T}$. We have $J_{\mathrm{IM}}T\gg t^2 \sim \Lambda^2$, so, to leading order we can neglect the dispersion in Fermion propagators. Then, there is nothing for the magnetic field to couple to, and consequently no magnetotransport. 

To illustrate this, let us compute the correlator of currents in perpendicular directions in real space on a square lattice. The uniform current operators are
\begin{align}
&I_x(\tau) \equiv \frac{1}{V^{1/2}}\sum_r I_{rx}(\tau) \equiv -\frac{it}{2V^{1/2}}\sum_{r;~i=1}^M c^\dg_{r+\hat{x},i}(\tau)c_{ri}(\tau) + \mathrm{h.c.}, \nn
&I_y(\tau) \equiv \frac{1}{V^{1/2}}\sum_r I_{ry}(\tau) \equiv -\frac{it}{2V^{1/2}}\sum_{r;~i=1}^M c^\dg_{r+\hat{y},i}(\tau)c_{ri}(\tau)e^{i\phi(r)} + \mathrm{h.c.},
\end{align}
where we have used a gauge with the magnetic vector potential $\mathbf{A}_r$ pointing along the $y$ direction, giving rise to the phase factors $e^{i\phi(r)}$ on bonds in the $y$ direction. The system volume in units of the unit cell volume is $V$. We then have
\begin{align}
&\mathcal{T}_\tau\langle I_x(\tau)I_y(\tau^\prime) \rangle = -M\frac{t^2}{4V}\sum_{rr^\prime}\Big[\mathcal{T}_\tau\langle c^\dg_{r+\hat{x}}(\tau)c_r(\tau)c^\dg_{r^\prime+\hat{y}}(\tau^\prime)c_r^\prime(\tau^\prime) e^{i\phi(r^\prime)}\rangle-\mathcal{T}_\tau\langle c^\dg_{r+\hat{x}}(\tau)c_r(\tau)c^\dg_{r^\prime}(\tau^\prime)c_{r^\prime+\hat{y}}(\tau^\prime) e^{-i\phi(r^\prime)}\rangle  \nn
&-\mathcal{T}_\tau\langle c^\dg_{r}(\tau)c_{r+\hat{x}}(\tau)c^\dg_{r^\prime+\hat{y}}(\tau^\prime)c_{r^\prime}(\tau^\prime) e^{i\phi(r^\prime)}\rangle+ \mathcal{T}_\tau\langle c^\dg_{r}(\tau)c_{r+\hat{x}}(\tau)c^\dg_{r^\prime}(\tau^\prime)c_{r^\prime+\hat{y}}(\tau^\prime) e^{-i\phi(r^\prime)}\rangle\Big],
\label{localII}
\end{align}
where we have dropped the sum over flavor indices in favor of a global factor of $M$, and $\mathcal{T}$ denotes time-ordering. To leading order in $t$, since the $c$ Green's functions are completely local,
\beq
\mathcal{T}_\tau\langle c_r(\tau)c^\dg_{r^\prime}(\tau^\prime)\rangle = \delta_{rr^\prime}G^c(\tau-\tau^\prime),
\eeq
none of the terms in (\ref{localII}) can be nonzero. Similarly, at $\mathcal{O}(t^2)$, there is no field-dependent correction to the $\langle I_x I_x\rangle$ correlator. 

Perturbing in $t$, in order for (\ref{localII}) to be nonzero, we need to insert hopping vertices in order to close the 4-point correlation functions of the $c$'s. To lowest order in $t$, this requires insertion of two hopping vertices into each of the 4-point correlation functions in (\ref{localII}), so that the connected contractions of $c$'s and $c^\dg$'s into local $c$ Green's functions go around a single plaquette of the lattice. Again, due to our choice of gauge, hopping vertices along bonds in the $y$ direction come with phase factors. But we obtain, as we should, a gauge-invariant answer for the connected part, which is of interest to us here (the electrons are negatively charged, and $\mathcal{B}$ is defined in terms of $B$ as in Sec.~\ref{mfltransport})
\beq
\langle I_xI_y \rangle(i\Omega_m) = -iM\sin(\mathcal{B})t^4T\sum_{\omega_n}[(G^c(i\omega_n))^3(G^c(i\omega_n+i\Omega_m)-G^c(i\omega_n-i\Omega_m))].
\label{localIIw}
\eeq
At $\mathcal{O}(t^4)$, vertex corrections from the coupling $g$ to this leading contribution vanish due to the non-correlation of $g$ between distinct lattice sites, i.e. $\ll g^r_{ijkl} g^{r^\prime}_{jilk} \gg = g^2 \delta_{rr^\prime}$. 
 
The DC Hall conductivity follows,
\begin{align}
&\sigma_H^{\mathrm{IM}} = -\lim_{\omega\rightarrow0}\frac{1}{i\omega}\left[\langle I_x I_y\rangle(i\Omega_m\rightarrow \omega+i0^+)-\langle I_x I_y\rangle(i\Omega_m\rightarrow 0+i0^+)\right] \nn
&=2M\sin(\mathcal{B})t^4\mathcal{P}\int_{-\infty}^{\infty}\frac{dE_1}{2\pi}\frac{dE_2}{2\pi}A_3^c(E_1)A^c(E_2)\frac{n_f(E_2)-n_f(E_1)}{(E_2-E_1)^2},
\end{align}
where $\mathcal{P}$ denotes the Cauchy principal value, and 
\beq
A_3^c(E_1) \equiv -2\mathrm{Im}[(G^c(i\omega_n\rightarrow E_1+i0^+))^3] = \mathrm{Im}\left[\frac{(i-1)(i+e^{2\pi\mathcal{E}_c})^3\cosh^{3/4}(2\pi\mathcal{E})}{2^{5/2}\pi^{9/4}J_{\mathrm{IM}}^{3/2}T^{3/2}(1+e^{4\pi\mathcal{E}_c})^{3/2}}\frac{\Gamma^3 \left(\frac{1}{4}-\frac{i(E_1-2\pi\mathcal{E}_cT)}{2 \pi  T}\right)}{\Gamma^3 \left(\frac{3}{4}-\frac{i(E_1-2\pi\mathcal{E}_cT)}{2 \pi  T}\right)}\right],
\eeq
is the spectral function of $(G^c(i\omega_n))^3$. If $\mathcal{E}_c=0$, then the Hall conductivity vanishes due to the evenness of the spectral functions $A^c$ and $A^c_3$. This corresponds to half-filling the square lattice, so this is expected. Scaling out $T$ and evaluating the integral numerically gives
\beq
\sigma_H^{\mathrm{IM}} = -M\sin(\mathcal{B})\frac{t^4\cosh(2\pi\mathcal{E})}{J_{\mathrm{IM}}^2T^2}\Xi^\mathrm{IM}_H(\mathcal{E}_c),
\eeq
where $\Xi^\mathrm{IM}_H(\mathcal{E}_c)$ is odd in $\mathcal{E}_c$, positive for positive $\mathcal{E}_c$, and vanishes when $\mathcal{E}_c = 0,\pm\infty$. This is a very small contribution regardless of $B$; the already small flux per unit cell $\mathcal{B}$ is further multiplied by a small parameter $t^4/(J_{\mathrm{IM}}^2T^2)$. Note that we consider $\cosh(2\pi\mathcal{E})$ to be $\mathcal{O}(1)$. If $|\mathcal{E}|$ is very large, then the conduction electrons do not scatter effectively off the islands, as discussed before, and our perturbative expansion in hopping is no longer valid, and in that case the system is once again described by the MFL. For the Hall conductivity to be comparable to the longitudinal conductivity $\sigma^{\mathrm{IM}}_0\sim t^2/(J_{\mathrm{IM}}T)$, we need $\sin(\mathcal{B})\sim J_{\mathrm{IM}} T/t^2 \gg 1$, which is not even mathematically possible. 

Similarly, the field-dependent correction to the $I_x$-$I_x$ correlator is
\beq
\Delta_B\left[\langle I_x I_x \rangle (i\Omega_m)\right] = -Mt^4\cos(\mathcal{B})T\sum_{\omega_n}(G^c(i\omega_n))^2(G^c(i\omega_n+i\Omega_m))^2,
\label{localIIxB}
\eeq
leading to the field-dependent correction to the longitudinal conductivity
\beq
\Delta_B[\sigma_L^{\mathrm{IM}}] = \frac{M}{8}\frac{t^4}{T}\cos(\mathcal{B})\int\frac{dE_1}{2\pi}A_2^c(E_1)\mathrm{sech}^2\left(\frac{E_1}{2T}\right),
\eeq
where 
\beq
A_2^c(E_1) \equiv -2\mathrm{Im}[(G^c(i\omega_n\rightarrow E_1+i0^+))^2] = -\mathrm{Im}\left[i\frac{(i+e^{2\pi\mathcal{E}_c})^2\cosh^{1/2}(2\pi\mathcal{E})}{2\pi^{3/2}J_{\mathrm{IM}}T(1+e^{4\pi\mathcal{E}_c})}\frac{\Gamma^2 \left(\frac{1}{4}-\frac{i(E_1-2\pi\mathcal{E}_cT)}{2 \pi  T}\right)}{\Gamma^2 \left(\frac{3}{4}-\frac{i(E_1-2\pi\mathcal{E}_cT)}{2 \pi  T}\right)}\right],
\eeq
is the spectral function of $(G^c(i\omega_n))^2$. Scaling out $T$ and evaluating the integral numerically gives
\beq
\Delta_B[\sigma_L^{\mathrm{IM}}] = M\frac{t^4\cosh(2\pi\mathcal{E})}{J_{\mathrm{IM}}^2T^2}\cos(\mathcal{B})\Xi^\mathrm{IM}_L(\mathcal{E}_c),
\eeq
where $\Xi^\mathrm{IM}_L(\mathcal{E}_c)$ is even in $\mathcal{E}_c$, positive, nonzero for $\mathcal{E}_c=0$, and vanishes as $\mathcal{E}_c \rightarrow \pm\infty$. The longitudinal conductivity is thus reduced when a field is applied, as is usually the case. 

\begin{figure}
\begin{center}
\includegraphics[height=1.7in]{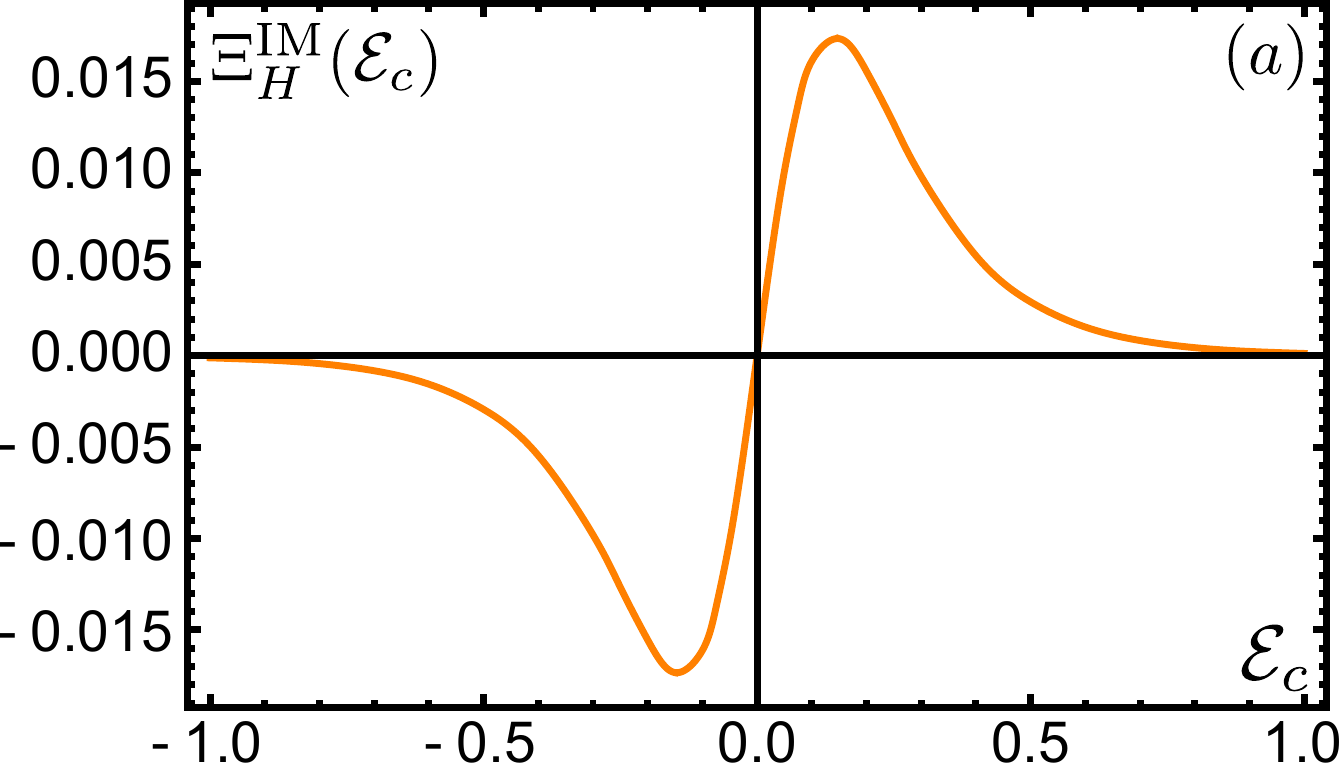}~~~~\includegraphics[height=1.7in]{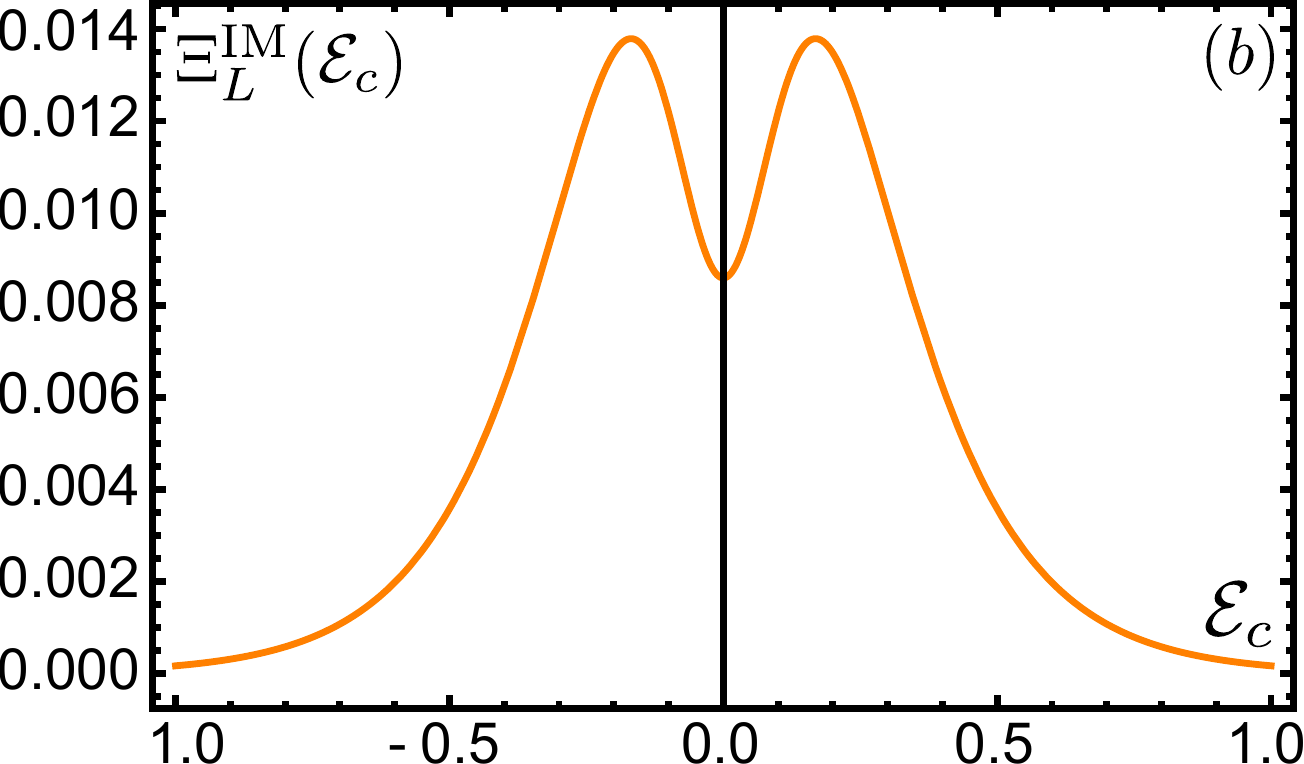}
\end{center}
\caption{Plots of (a) $\Xi^{\mathrm{IM}}_{H}(\mathcal{E}_c)$ and (b) $\Xi^{\mathrm{IM}}_{L}(\mathcal{E}_c)$. Both functions vanish in the limits of the fully filled and empty lattice ($\mathcal{E}_c = \mp \infty$ respectively), as they should.}
\label{Xiplots}
\end{figure}

It is similarly thus not possible to get a field-dependent correction to $\sigma^{\mathrm{IM}}_L$ that is comparable to its zero-field value. Thus we shall no more consider the IM regime for studying magnetotransport, as there is no qualitative difference between the regimes of `large' and small $B$ unlike in the MFL regime. For completeness, the plots of $\Xi^{\mathrm{IM}}_{H,L}(\mathcal{E}_c)$ are shown in Fig.~\ref{Xiplots}.

Before we close this section, let us comment on the controllability of the hopping expansion used to compute the nonzero field-dependent conductivity corrections. Clearly, this hopping expansion must break down when $t$ is large enough, as the MFL has a very different conductivity tensor. Going from (\ref{localII}) to (\ref{localIIw}) and (\ref{localIIxB}), we only kept those $r^\prime$ relative to $r$ that resulted in $\mathcal{O}(t^4)$ corrections for the shortest closed paths from $r$ to $r^\prime$ and back. For arbitrary $r^\prime$, one can draw infinitely many paths that go from $r$ to $r^\prime$ and back. These paths may also intersect themselves in general. For a path length $l$, there are $< 4^l$ paths for large $l$, as at each step, one has $4$ choices of direction, and not all possibilities will result in a formation of the closed path from $r$ to $r^\prime$ and back. Each step involves mulitplying an additional local Green's function and factor of $t$, or roughly a factor of $\sim t/(J_{\mathrm{IM}}T)^{1/2}\ll 1$ into the amplitude. Therefore, the total weight of paths of length $l$ should be $< (4t/(J_\mathrm{IM}T)^{1/2})^l$. The total weight of all paths between $r,r^\prime$ then is $< \sum_{l=l_{\mathrm{min}}}^\infty (4t/(J_\mathrm{IM}T)^{1/2})^l =  (4t/(J_\mathrm{IM}T)^{1/2})^{l_\mathrm{min}}/(1-4t/(J_\mathrm{IM}T)^{1/2})$, where $l_\mathrm{min}$ is the length of the shortest closed path between $r,r^\prime$, which scales as the lattice distance between $r,r^\prime$. Thus, for $t/(J_{\mathrm{IM}}T)^{1/2}\ll 1$, the expansion is well behaved: as $r^\prime$ gets further away from $r$, the terms are exponentially suppressed in the distance between $r$ and $r^\prime$, whereas the number of $r^\prime$'s a given distance away from $r$ grows only linearly in that distance in two dimensions. Unsurprisingly, this is just the condition $T\gg T_\mathrm{inc}$ we obtained earlier for the crossover into the IM regime.
 
\section{Macroscopic transport via Effective-medium/Random-resistor theory}
\label{EMARRN}

\change{We now return to the MFL with $B/T$ scaling that was described in Section~\ref{mfltransport}. We will show here that adding macroscopic
disorder leads to a linear-in-$B$ magnetoresistance at large $B$, while preserving the $B/T$ scaling. We will treat the inhomogeneity in a classical transport
framework. The quantum computation in Section~\ref{mfltransport} is used to compute a local $\sigma_{xx}$ and $\sigma_{xy}$, which is then in put
into a computation of global transport in a disordered sample by composing resistivities using Ohm's and Kirchhoff's laws.}

\subsection{Setup}

We seek to understand the effects of additional macroscopic disorder on the transport of charge in the MFL at `large' magnetic fields $B$, in two spatial dimensions. This additional macroscopic disorder leads to the variation of the local conductivity tensor $\mathbf{\sigma}(\mathbf{x})$ across the sample. Since the conduction electrons in our model interact with valence electrons in the islands through a non-translationally invariant interaction microscopically, the Navier-Stokes equation of hydrodynamics that describes dynamics of a nearly-conserved macroscopic momentum~\cite{Hartnoll2016} is not applicable to us, since this requires microscopic equilibriation of the electron fluid through {\it momentum-conserving} interactions (the effects of weak disorder on the magnetoresistance of a generic electron fluid with macroscopic momentum were studied in Ref.~\onlinecite{Patel2017}; they did not find any regimes of linear magnetoresistance, instead finding that the magnetoresistance was quadratic with a prefactor controlled by the fluid viscosity). Thus, at the coarse-grained level, we just have the equation for charge conservation, and Ohm's law
\beq
\nabla\cdot\mathbf{I}(\mathbf{x}) = 0,~~\mathbf{I}(\mathbf{x}) = \mathbf{\sigma}(\mathbf{x})\cdot \mathbf{E}(\mathbf{x}),~~\mathbf{E}(\mathbf{x}) = -\nabla\Phi(\mathbf{x}).
\label{basic}
\eeq
The effective local electric field $\mathbf{E}(\mathbf{x})$ (which includes the effects of Coulomb potentials generated due to charge inhomogeneities~\cite{Lucas2016}) fluctuates spatially due to the macroscopic disorder, but equals an applied external electric field $\mathbf{E}_0=\langle\mathbf{E}(\mathbf{x})\rangle \equiv \frac{1}{V}\int d^2\mathbf{x}~\mathbf{E}(\mathbf{x})$ on spatial average. We define the global conductivity tensor $\mathbf{\sigma}^e$ through the relation $\langle\mathbf{I}(\mathbf{x})\rangle = \mathbf{\sigma}^e \cdot \mathbf{E}_0$, and parameterize the deviation $\mathbf{\sigma}(\mathbf{x}) -\mathbf{\sigma}^e = \delta\mathbf{\sigma}(\mathbf{x})$. The condition $\langle \mathbf{I}(\mathbf{x})-\langle \mathbf{I}(\mathbf{x}) \rangle\rangle = 0$ then gives $\langle \mathbf{\chi}(\mathbf{x})\cdot \mathbf{E}_0 \equiv \delta\mathbf{\sigma}(\mathbf{x})\cdot \mathbf{E}(\mathbf{x})\rangle = 0$. 

Following Ref.~\onlinecite{Stroud1975}, without making any additional approximations, the solution of these equations can be formally cast in the form
\beq
\Phi(\mathbf{x}) = -\mathbf{E}_0\cdot\mathbf{x} + \int d^2\mathbf{x}^\prime~\mathcal{G}(\mathbf{x},\mathbf{x}^\prime)\nabla^\prime\cdot(\delta\mathbf{\sigma}(\mathbf{x}^\prime)\cdot\nabla^\prime\Phi(\mathbf{x}^\prime)),
\eeq
where the Green's function satisfies $\nabla\cdot(\mathbf{\sigma}^e\cdot\nabla\mathcal{G}(\mathbf{x},\mathbf{x}^\prime))=-\delta(\mathbf{x}-\mathbf{x}^\prime)$, $\mathcal{G}(\mathbf{x},\mathbf{x}^\prime) = \mathcal{G}(\mathbf{x}^\prime,\mathbf{x})$, and $G(\mathbf{x},\mathbf{x}^\prime\in\partial V)=0$, for the system boundary $\partial V$, which we take to infinity. Taking a gradient on both sides, we get
\begin{align}
&\mathbf{E}(\mathbf{x}) = \mathbf{E}_0 - \int d^2\mathbf{x}^\prime~[(\delta\mathbf{\sigma}(\mathbf{x}^\prime)\cdot\mathbf{E}(\mathbf{x}^\prime))\cdot\nabla^\prime]\cdot\nabla\mathcal{G}(\mathbf{x},\mathbf{x}^\prime),~~\mathrm{or} \nn
&\mathbf{\chi}(\mathbf{x}) = \delta\mathbf{\sigma}(\mathbf{x})- \delta\mathbf{\sigma}(\mathbf{x})\cdot\int d^2\mathbf{x}^\prime~\mathcal{K}(\mathbf{x},\mathbf{x}^\prime)\cdot\mathbf{\chi}(\mathbf{x}^\prime),
\end{align}
where the second line follows from the first by left-multiplying both sides by $\delta\mathbf{\sigma}(\mathbf{x})$, and then demanding that it hold for any $\mathbf{E}_0$, and $\mathcal{K}_{ij}(\mathbf{x},\mathbf{x}^\prime) = \partial_i\partial^\prime_j\mathcal{G}(\mathbf{x},\mathbf{x}^\prime)$.

We now assume that the disorder divides the sample into macroscopic domains whose size is much smaller than the sample size, but much bigger than the smaller of the electron mean-free path and electron cyclotron radius, and the tensors $\mathbf{\chi}$ and $\delta\mathbf{\sigma}$ take on constant values in a given domain. For a given domain $p$, we can write
\beq
\mathbf{\chi}^p = \delta\mathbf{\sigma}^p- \delta\mathbf{\sigma}^p\cdot\int_p d^2\mathbf{x}^\prime~\mathcal{K}(\mathbf{x}\in p,\mathbf{x}^\prime)\cdot\mathbf{\chi}^p - \delta\mathbf{\sigma}^p\cdot\sum_{p^\prime\neq p}\cdot\int_{p^\prime} d^2\mathbf{x}^\prime~\mathcal{K}(\mathbf{x}\in p,\mathbf{x}^\prime)\cdot\mathbf{\chi}^{p^\prime}.
\label{EMApre}
\eeq
For the second integral over domains other than the given domain, we replace $\mathbf{\chi}^n$ with its spatial average $\langle \mathbf{\chi}\rangle$. This is the `effective-medium' approximation~\cite{Stroud1975}: The equivalent conductivity of each domain is controlled in part by a `mean-field' of domains surrounding it. However, since our conventions are set up so that $\langle \mathbf{\chi}\rangle = 0$, this second term drops out. Then, spatially averaging both sides, we obtain 
\beq
\sum_p V^p \mathbf{\chi}^p = 0~~\Rightarrow~~\sum_p V^p(\mathbb{I}+\delta\mathbf{\sigma}^p\cdot\mathcal{M}^p)^{-1}\cdot\delta{\mathbf{\sigma}}^p = 0,
\label{EMA}
\eeq
where $V^p$ is the volume fraction of domain $p$ and $\mathcal{M}^p_{ij} = \oint_{\partial^\prime p}\partial_i\mathcal{G}(\mathbf{x},\mathbf{x}^\prime)\hat{n}^{\prime p}_j$, where the integral is over the primed coordinate, and $\hat{\mathbf{n}}^{\prime p}$ is the outward-pointing unit normal vector on the boundary of $p$, varying with the primed coordinate. 

If the local conductivity tensor $\mathbf{\sigma}(\mathbf{x})$ is known in all domains, (\ref{EMA}) can then be solved for $\mathbf{\sigma}^e$. In our two-dimensional electron problem, we expect  $\mathbf{\sigma}^e_{ij} = \delta_{ij}\sigma^e_L-\epsilon_{ij}\sigma^e_H$, where $\sigma^e_L$ is even in $B$ and $\sigma^e_H$ is odd in $B$ because of Onsager reciprocity, so we obtain the Green's function $\mathcal{G}(\mathbf{x},\mathbf{x}^\prime) = -\ln(|\mathbf{x}-\mathbf{x}^\prime|0^+)/(2\pi\sigma^e_L)$. Then, for circular domains, $\mathcal{M}^p_{ij} = \delta_{ij}/(2\sigma^e_L)$ is indeed independent of $\mathbf{x}$. This makes (\ref{EMApre}) and (\ref{EMA}) self-consistent~\cite{Stroud1975}. For other domain shapes, there are corrections when $\mathbf{x}$ is near the domain boundary.

For an analytically solvable toy model, we assume that the $\mathbf{\sigma}(\mathbf{x})$ can take either of two possible values $\mathbf{\sigma}^a$ and $\mathbf{\sigma}^b$ in circular domains that are spatially randomly distributed over the sample~\cite{Guttal2005,Dykhne1971} (Fig~\ref{Redpillgreenpill}a). As far as the asymptotic low and high-field magnetoresistance goes, this already yields the same qualitative behavior at large and small fields as a more complicated model with a distribution of different types of domains~\cite{Ramakrishnan2017}. Furthermore, the `mean-field' like effective-medium approximation has also been shown to produce results for the magnetoresistance equivalent to exact numerical solutions of (\ref{basic}) in random-resistor network models~\cite{Ramakrishnan2017,Parish2005,Parish2003}. In the simplified two-type scenario (\ref{EMA}) then simplifies to~\cite{Guttal2005}
\beq
V^a\left(\mathbb{I}+\frac{\mathbf{\sigma}^a-\mathbf{\sigma}^e}{2\sigma^e_L}\right)^{-1}\cdot(\mathbf{\sigma}^a-\mathbf{\sigma}^e) + (1-V^a)\left(\mathbb{I}+\frac{\mathbf{\sigma}^b-\mathbf{\sigma}^e}{2\sigma^e_L}\right)^{-1}\cdot(\mathbf{\sigma}^b-\mathbf{\sigma}^e) = 0.
\label{EMAsimple}
\eeq
If $V^a=1/2$, this yields an unsaturating high-field linear magnetoresistance~\cite{Guttal2005}. For the model with a distribution of domains, the equivalent condition is that the distribution is symmetric about its mean~\cite{Ramakrishnan2017}. For $V^a$ detuned from $1/2$, the magnetoresistance saturates, but there is an intermediate regime of fields in which the magnetoresistance is approximately linear, and the saturation field becomes arbitrarily large as $V^a$ approaches $1/2$~\cite{Guttal2005}. The rough reasoning behind the saturation appears to be that, if one type of domain is far more common than the other, the current flowing through the sample mainly finds paths involving only one type of domain, and hence the global magnetoresistance behaves like that of a single domain, which saturates at high fields~\cite{Parish2005}. We will do our analysis with the symmetric distribution $V^a = 1- V^a = 1/2$. 

A physical picture for the high-field linear magnetoresistance was provided in Ref.~\onlinecite{Parish2003}, and involves the contribution of the local Hall resistance (which is linear in $B$) to the global longitudinal resistance due to the distortion in current paths arising from spatial fluctuations of the local Hall resistance: In a uniform sample, charge accumulation at the edges of the sample parallel to the applied electric field produces a global Hall electric field perpendicular to the applied electric field that cancels out Hall currents throughout the sample. On the other hand, if the sample has a disordered local conductivity tensor, the global Hall electric field no longer cancels out local Hall currents throughout the sample. Thus, the global longitudinal resistance becomes dependent on the local Hall resistances.       

\subsection{Application}

We note that in (\ref{mfllcmag}), the $\mathrm{sech}$ is strongly peaked near $E_1=0$, whereas for a finite temperature, $\mathrm{Im}[\Sigma^c_R(E_1)]$ does not vary drastically with $E_1$ near $E_1=0$ over the range which the $\mathrm{sech}$ is appreciable. We can thus replace $\mathrm{Im}[\Sigma^c_R(E_1)]$ with $\gamma/2$ from (\ref{dumprate}). Regardless of this approximation, we note from (\ref{mfllcmag}) that $\sigma^{\mathrm{MFL}}_L \sim T/B^2$ and $\sigma^{\mathrm{MFL}}_H \sim 1/B$ at large $B$, which is what the effective-medium theory needs to produce linear magnetoresistance at large $B$. This asymptotic scaling holds even if we had multiple MFL bands, thus adding their conductivity tensors to get the appropriate local conductivity tensor. 

\begin{figure}
\begin{center}
\includegraphics[height=2.5in]{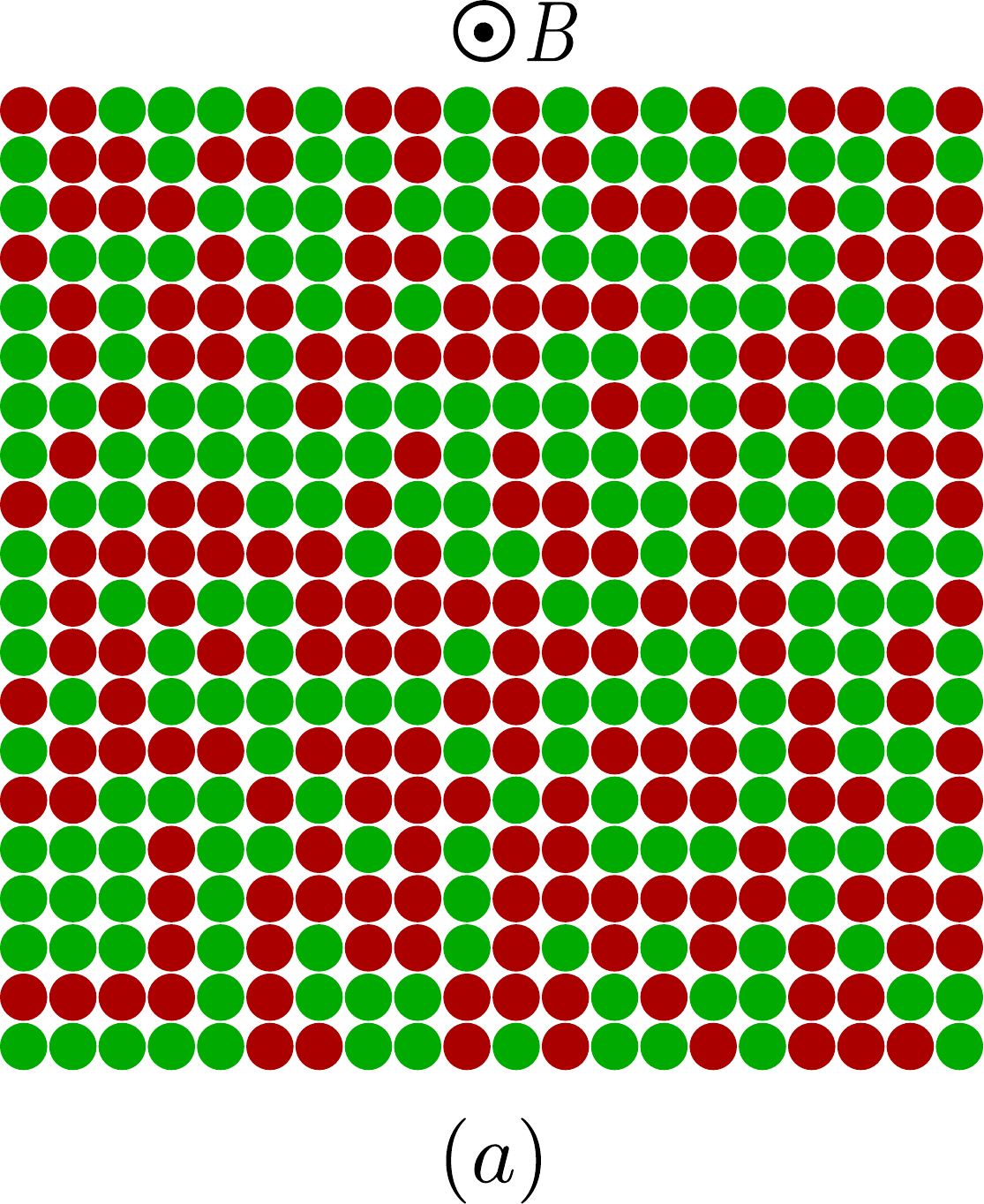}~~~~~~~~~~\includegraphics[height=2.5in]{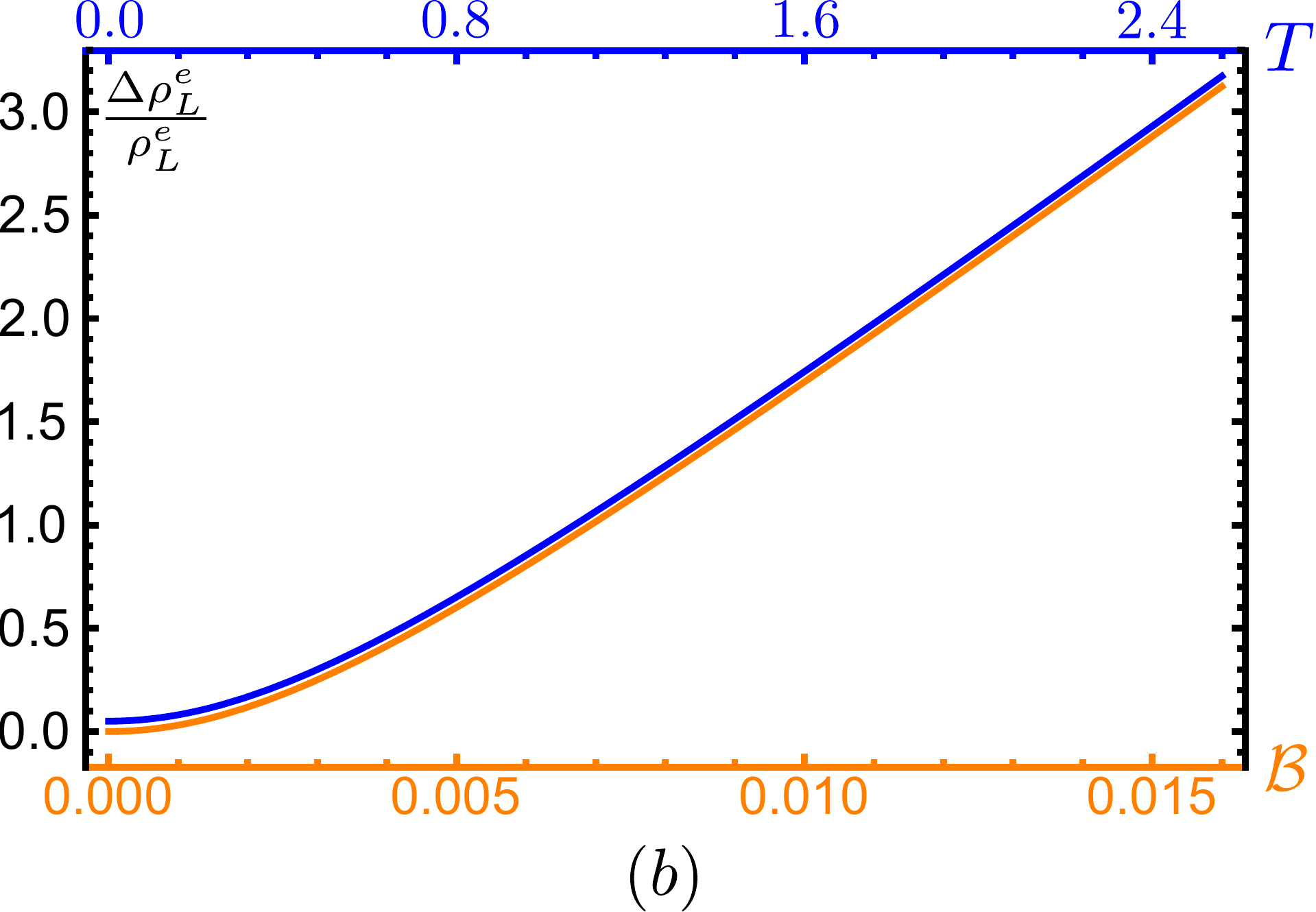}
\end{center}
\caption{(a) A cartoon of a two-dimensional sample with a random distribution of approximately equal fractions of two types of domains, for which an exact analytic solution of the effective-medium equations for magnetotransport is possible. The magnetic field $B$ points out of the plane of the sample. (b) Plots of the normalized change in global longitudinal resistance due to dimensionless magnetic field $\mathcal{B}$ (orange) and due to temperature $T$ (blue), obtained from (\ref{mtp2}). We use $E^F_b/E^F_a=0.8$ and $\gamma_b/\gamma_a=0.8$. The dimensionless magnetic field $\mathcal{B}$ is the flux per unit cell $Ba^2$ in units of $\hbar/e$ (\ref{BcalB}). We use $m=0.005\sim 1/E^F_{a,b}$. The orange ($\mathcal{B}$) curve is evaluated at $T=1.0$ and $\gamma_a=0.1$ and the blue ($T$) curve is evaluated at $\mathcal{B}=0.0025$ and $\gamma_a=0.1 T$. The curves are slightly offset for visualization, but actually lie on top of each other, demonstrating a scaling between magnetic field and temperature. Both the $\mathcal{B}$ and $T$ dependencies are quadratic at small fields \change{or} temperatures and cross over to linear at large fields \change{or} temperatures.}
\label{Redpillgreenpill}
\end{figure}

We thus input the following conductivity tensors into the effective-medium calculation (we take the band mass $m=k_F/v_F$ to be the same in both types of electron-like domains $a$ and $b$):
\beq
\sigma^{a,b}_{ij} = \frac{\sigma^{\mathrm{MFL}}_{0a,b}}{1+\mathcal{B}^2/(m\gamma_{a,b})^2}\left(\delta_{ij} + \epsilon_{ij} \frac{\mathcal{B}}{m\gamma_{a,b}}\right). 
\label{mflsigma}
\eeq
The scattering rate $\gamma$ can fluctuate across domains due to fluctuations in $g$, induced by fluctuations in the densities of islands, and the base conductivity $\sigma^{\mathrm{MFL}}_{0a,b}$ can fluctuate across domains due to fluctuations in both $g$ and in the electron density. Then, solving~(\ref{EMAsimple}) for $V^a=1-V^a=1/2$,  we get the global longitudinal and Hall resistances respectively,
\begin{align}
&\rho^e_L \equiv \frac{\sigma^e_L}{\sigma^{e2}_L+\sigma^{e2}_H} = \frac{\sqrt{(\mathcal{B}/m)^2\left(\gamma_a \sigma^{\mathrm{MFL}}_{0a}-\gamma _b \sigma^{\mathrm{MFL}}_{0b}\right)^2+\gamma _a^2 \gamma _b^2 \left(\sigma^{\mathrm{MFL}}_{0a}+\sigma^{\mathrm{MFL}}_{0b}\right)^2}}{\gamma _a \gamma _b (\sigma^{\mathrm{MFL}}_{0a}\sigma^{\mathrm{MFL}}_{0b})^{1/2} \left(\sigma^{\mathrm{MFL}}_{0a}+\sigma^{\mathrm{MFL}}_{0b}\right)}, \nn
&\rho^e_H \equiv -\frac{\sigma^e_H/\mathcal{B}}{\sigma^{e2}_L+\sigma^{e2}_H} = \frac{\gamma _a+\gamma _b}{m \gamma _a \gamma _b \left(\sigma^{\mathrm{MFL}}_{0a}+\sigma^{\mathrm{MFL}}_{0b}\right)}.
\label{mtp1}
\end{align}
The magnetoresistance $\rho^e_L(\mathcal{B})-\rho^e_L(0)$ is thus linear as promised at high fields, and is quadratic at low fields. 
 
Considering the isotropic parabolic dispersion $\varepsilon_k = k^2/(2m)-\Lambda/2$, and using (\ref{s0MFL}), (\ref{dumprate}), and $\nu(0)=m$, we can write $\sigma^{\mathrm{MFL}}_{0a,b}= M w_\sigma E_F^{a,b}/\gamma_{a,b}$, where $w_\sigma=0.135689$ and $E_F^{a,b} = m v_{Fa,b}^2/2$ are the Fermi energies. We can then rewrite (\ref{mtp1}) as
\beq
w_\sigma\rho^e_L = \frac{\left(\gamma_a^2+\left(\frac{\mathcal{B}}{m}\right)^2\frac{\left(1-E^F_b/E^F_a\right)^2}{\left(\gamma_b/\gamma_a+E^F_b/E^F_a\right)^2}\right)^{1/2}}{M(\gamma_a/\gamma_b)^{1/2}(E^F_aE^F_b)^{1/2}},~~w_\sigma\rho^e_H = \frac{(1+\gamma_b/\gamma_a)}{M mE^F_a(E^F_b/E^F_a+\gamma_b/\gamma_a)}.
\label{mtp2}
\eeq
Plots of the normalized change in $\rho^e_L$ due to $\mathcal{B}$ and $T$ are shown in Fig.~\ref{Redpillgreenpill}b. This simplified model with two types of domains thus leads to a global longitudinal resistance that adds $T$ and $B$ in quadrature\footnote{Holographic realizations of a variety of magnetoresistance scalings, including quadrature, were found in Ref.~\onlinecite{Kiritsis2017}.}, as seen in the experiment of Ref.~\onlinecite{Hayes2016}. A continuous gaussian distribution of electron densities across the domains will also yield a qualitatively similar scaling function to the above quadrature function~\cite{Ramakrishnan2017}. In general, the zero-field linear-in-$T$ and high-field linear-in-$B$ behavior (as well as the scaling between $B$ and $T$) will emerge universally from such resistor-network models, but the interpolation between the two regimes is sensitive to the distribution of the local conductivity tensors.   

The Hall resistance is $\rho^e_H$ is sensitive to the disorder distribution and thus is not trivially controlled by the average carrier density $\propto Mm(E^F_b+E^F_a)/2$ even for the isotropic Fermi surfaces we consider, unless $\gamma_a=\gamma_b$. In this simplified version of the problem, $\rho^e_H$ is independent of temperature. However, we expect that more complicated disorder distributions generically give rise to some temperature dependence of $\rho^e_H$, which would depend on the disorder distribution even at a qualitative level. A detailed analysis of such effects is beyond the scope of the present work, and will be considered in the future. 

Since $\gamma_{a,b} \propto T$, the crossover from quadratic to linear magnetoresistance occurs at a field scale proportional to temperature. Additionally, if we use the `Planckian' choice of parameters, and if the disorder distribution is such that $|1-E^F_b/E^F_a|/(\gamma_b/\gamma_a+E^F_b/E^F_a)$ is an $\mathcal{O}(1)$ number, the crossover occurs at a field scale given by $\mu_B B \sim k_B T$, as discussed at the end of Sec.~\ref{mfltransport}. While this is most definitely a fine-tuned situation, and would require substantial variation in the charge densities between domains, it is within the scope of our theory. Alternatively, if $\gamma_a(\gamma_b/\gamma_a+E^F_b/E^F_a)/(k_B T |1-E^F_b/E^F_a|)$ is an $\mathcal{O}(1)$ quantity (but $\gamma_a\propto T$ is much smaller than $k_B T$), then $\rho^e_L$ can still be controlled by the approximate scaling function $\sqrt{1+(\mu_B B)^2/(k_B T)^2}$ for much smaller variations in the charge densities between domains.  

The effective-medium theory is applicable when the domain sizes are much greater than the smaller of the electron mean free path and electron cyclotron radius in a single domain. At low temperatures and weak fields, electrons can move through a domain without significant loss or deflection of momentum, and the effects of scattering off the boundaries between domains then become important, adding a temperature-independent residual resistivity to the result of the above computation.  

In our analysis, we have neglected the effects of the feedback of heat currents on charge transport. In general, one would have an additional analogous set of equations to (\ref{basic}) for heat currents and temperature gradients in place of charge currents and electric fields. Since there is no concept of bulk fluid motion due to translational symmetry breaking at the microscopic level, the equations for heat currents and charge currents would only be coupled if the local thermoelectric tensor $\mathbf{\alpha}(\mathbf{x})$ were nonzero. However, in the MFL, with $T\ll E^F_{a,b}$, $\mathbf{\alpha}(\mathbf{x})$ is negligible as discussed in Sec.~\ref{mfltransport}, and our decoupled analysis of charge currents is hence still applicable. Somewhere in the crossover region between the MFL and the IM, a regime may exist where both $\mathbf{\alpha}(\mathbf{x})$ and the effects of magnetic fields on the local conductivity tensors are simultaneously significant, and there may be a significant feedback of thermoelectric effects on the charge magnetotransport. We leave a detailed study of such effects for future work. 

\section{Discussion}
\label{discuss}

The strange metal phases of the cuprate and pnictide high-$T_c$ superconductors occur at finite dopings, and consequently display significant amounts of disorder. Experimentally, there is direct evidence for disorder at (i) microscopic levels, due to irregular placements of dopant atoms~\cite{McElroy2005}, and (ii) meso- and macroscopic levels, due to a variety of factors ranging from crystalline imperfections to charge puddles caused by impurities and non-isovalent dopants~\cite{Hanaguri2007,Kohsaka2012}. Additionally, due to these materials being layered, with relatively poor interlayer conductivities, imperfections in a layer may further induce heterogeneities in the charge distributions of adjacent layers through Coulomb forces.  

We have attempted to paint an impressionist picture of transport and magnetotransport in a strange metal by developing a solvable model that incorporates disorder at both microscopic and macroscopic levels. At the microscopic level, we built off remarkable recent developments~\cite{Gu2017,Sachdev2017,Balents2017,Zhang2017,Haldar2017,McGreevy2017} in realizing \rchange{solvable} field-theoretic  descriptions of extended non-Fermi liquid phases using SYK models. \rchange{These models couple together SYK quantum islands without quasiparticle excitations, and show how this can lead to non-Fermi liquid transport in an extended finite-dimensional phase. In our model we} locally and randomly couple mobile conduction electrons to immobile \rchange{quantum} islands described by SYK models in a particular way. \rchange{In this manner we realized a disordered marginal Fermi liquid} (MFL) phase \change{at low temperatures} with a linear-in-$T$ resistivity, \change{and an identifiable Fermi surface.} We determined the two-point functions, conductivities, and magnetotransport properties of this phase exactly in two spatial dimensions, finding a scaling between magnetic field and temperature in the conductivity tensor. Additionally, we showed that nearly-local `incoherent-metal' (IM) phases, 
\change{with no
identifiable Fermi surface}, are also realized in our model \change{at higher temperatures} in certain parameter regimes; \change{these} IMs can also have linear-in-$T$ resistivities, but have very weak effects of magnetic fields on their charge transport properties, making them unlikely candidates for a description of the strange metals seen in experiments at lower temperatures, \change{ which is where the large linear-in-$B$ magnetoresistances are also observed}. However, the IMs may still be the correct concept at high temperatures, due to strong bad-metallic behavior displayed through their large resistivities, \change{as is seen in experiments. It should also be noted that the large linear magnetoresistances are {\it not} observed in experiments performed at high temperatures where the system is a bad metal, with a zero-field resistivity much larger than the quantum unit $h/e^2$~\cite{Hayes2016,Giraldo2017}, which is consistent with the behavior of an IM.} 

\change{While the MFL regime of our model does indeed have a linear-in-$T$ resistivity, and also a $B/T$ scaling at approximately the 
observed $B$ scale, it yields a magnetoresistance which saturates at large $B$. To obtain a non-saturating magnetoresistance, we argued for the importance
of macroscopic disorder in the MFL regime. To model such effects,} we applied the effective-medium approximation to a sample containing domains of our disordered linear-in-$T$ MFLs with varying \change{electron} densities. While the effective-medium approximation is a mean-field theory at the level of Kirchhoff's and Ohm's laws for current flow, it has shown to be equivalent to exact numerical simulations of random-resistor networks for magnetotransport~\cite{Ramakrishnan2017}, and has also had remarkable successes in describing experimentally observed magnetoresistances in other two-dimensional disordered materials~\cite{Ramakrishnan2017,Ping2014,Ramakrishnan2015}. For certain simplified disorder distributions, the effective-medium equations for magnetotransport are analytically solvable. These exactly solvable equations yield, in our case, a magnetoresistance that is quadratic in field at low fields, crosses over to linear in field at high fields, and is controlled by a scaling function between field and temperature, as seen in recent experiments on the pnictide and cuprate strange metals~\cite{Hayes2016,Giraldo2017}.

On the experimental front, the anomalous high-field linear magnetoresistance in the cuprate and pnictide strange metals is already known to be dependent on the component of the magnetic field perpendicular to the sample plane~\cite{AnalytisPrivate}, a feature that our model reproduces, since it is based on orbital effects of the magnetic field on charge transport. Furthermore, a strong linear component of the high-field magnetoresistance is seen even away from the critical doping at which the zero-field resistance is almost exactly linear-in-$T$~\cite{Hayes2016,Giraldo2017}. The disorder based mechanism considered by us would be consistent with this observation, as the zero-field linear-in-$T$ behavior is not a prerequisite for high-field disorder-induced linear magnetoresistance; all that is required is that the local conductivity tensor behaves like (\ref{mflsigma}) as a function of magnetic field.

On the theoretical front, we have been able to analytically calculate non-trivial magnetotransport properties in a somewhat contrived, but solvable, model of a disordered non-Fermi liquid. 
Studies along the lines of Refs.~\onlinecite{MSB89,BF92,senthilLM} could
show how such models emerge naturally as effective theories of realistic, disordered, single-band Hubbard models.
We hope that our study motivates further investigations into the interplay of disorder and strong interactions in the transport properties of the strange metal phases of the pnictides and cuprates.       

\section*{Acknowledgements}

We thank J. G. Analytis, L. Balents, J. C. S\'eamus Davis, E.-A. Kim, A. Lucas and B. Ramshaw for useful discussions. AAP was supported by the NSF Grant PHY-1125915 via a KITP Graduate Fellowship. This research was also supported by the NSF Grant DMR-1664842,  by the MURI grant W911NF-14-1-0003 from ARO, and by funds provided by the U.S. Department of Energy (DOE) under cooperative research agreement DE-SC0009919. Research at Perimeter Institute is supported by the Government of Canada through Industry Canada and by the Province of Ontario through the Ministry of Research and Innovation. SS also acknowledges support from Cenovus Energy at Perimeter Institute and from the Hanna Visiting Professor program at Stanford University.

As this work was nearing completion, we learned of related but independent work by Chowdhury et. al. on realizing translationally-invariant microscopic models of non-Fermi liquids using SYK models~\cite{Chowdhury2018}.

\appendix

\section{Effects of `Pair-hopping' and bilinear terms on the marginal-Fermi liquid}
\label{pairhop}

We consider the effects of the `pair-hopping' term~(\ref{hph}) on the MFL as $T\rightarrow0$. With the Hamiltonian given by~(\ref{hph}), the Dyson equations are given by
\begin{align}
&\Sigma(\tau) = - J^2 G^2(\tau)G(-\tau) - \frac{M}{N} g^2 G(\tau)G^c(\tau)G^c(-\tau)  - \frac{M}{N} \eta^2 G(-\tau)(G^c(\tau))^2, \nn
&G(i\omega_n) = \frac{1}{i\omega_n + \mu - \Sigma(i\omega_n)}, \nn
&\Sigma^c(\tau) = -g^2G^c(\tau)G(\tau)G(-\tau) -\eta^2G^c(-\tau)(G(\tau))^2, \nn
&G^c(i\omega_n) = \int \frac{d^dk}{(2\pi)^d} \frac{1}{i\omega_n - \epsilon_k + \mu- \Sigma^c(i\omega_n)}.
\end{align}
If $\mu=0$, the exact relations $G(\tau)=-G(-\tau)$ and $G^c(\tau)=-G^c(-\tau)$ imply that the only effect of the pair-hopping term on the physics considered in the main text in all regimes is just a redefinition of $g$, with $g\rightarrow(g^2+\eta^2)^{1/2}$.

As long as the bandwidth is large, i.e. $t\gg g,\eta,J$,~(\ref{gc0}) is still valid. Following the same procedure as we did in Sec.~\ref{infiniband}, and using $G(\tau)$ given by~(\ref{sykconf0}), we obtain
\begin{align}
&\Sigma^c(i\omega_n\rightarrow0) = \frac{ig^2\nu(0)}{2J\cosh^{1/2}(2\pi\mathcal{E})\pi^{3/2}}\omega_n\ln\left(\frac{|\omega_n|e^{\gamma_E-1}}{J}\right) \nn
&+ \frac{\eta^2\nu(0)\cosh^{1/2}(2\pi\mathcal{E})}{2\pi^{3/2}}\left(i\frac{\omega_n}{J}\ln\left(\frac{|\omega_n|e^{\gamma_E-1}}{J}\right)-\tanh(2\pi\mathcal{E})\right) + \mathcal{O}(\omega_n).
\end{align}
This is clearly a marginal-Fermi liquid with an additional chemical potential correction
\beq
\delta \mu = \frac{\eta^2\nu(0)\cosh^{1/2}(2\pi\mathcal{E})}{2\pi^{3/2}}\tanh(2\pi\mathcal{E}) \ll \Lambda,
\eeq
which leads to a harmless small change in the size of the conduction electron Fermi surface, as the numbers of $c$ and $f$ electrons are no longer independently conserved (but their sum is conserved). 

There is also a back-reaction to the SYK islands
\begin{align}
&\tilde{\Sigma}(\tau) = -\frac{M}{N}g^2G(\tau)G^c(\tau)G^c(-\tau) -\frac{M}{N}\eta^2G(-\tau)(G^c(\tau))^2, \\
&\tilde{\Sigma}(i\omega_n\rightarrow0) = \frac{Mg^2(\nu(0))^2J\sinh(\pi\mathcal{E})}{3\sqrt{2}N\pi^{9/4}\cosh^{1/4}(2\pi\mathcal{E})} + \mathcal{O}(\omega_n),
\end{align}
which is again a chemical potential correction plus irrelevant frequency-dependent corrections. This chemical potential correction actually changes $\mathcal{E}$, which is no longer a conserved quantity, and is determined by the condition $\mathrm{Re}[\Sigma(i\omega_n\rightarrow0)]=\mu+\delta\mu$.

We also briefly discuss qualitatively the effects of certain fermion bilinears in (\ref{ham}). Terms bilinear in the $f$'s destroy their SYK behavior and non-zero entropy as $T\rightarrow0$. The $c$'s then scatter off essentially non-interacting random-matrix islands, with $G(i\omega_n)\sim i\mathrm{sgn}(\omega_n)$. This leads to $\mathrm{Im}[\Sigma^c_R(0)]\sim T^2$, and the $c$'s hence realize a weakly-interacting disordered Fermi liquid as $T\rightarrow0$. However, if the coefficients of the $f$-bilinears are small, then their SYK behavior is restored for temperatures larger than a small energy scale $E_c$~\cite{Balents2017}. Hence, the marginal-Fermi liquid behavior of the $c$'s is also restored for $T>E_c$.

The effects of bilinears which hybridize $c$'s and $f$'s (such as $c^\dg f$) were disussed in Ref.~\cite{McGreevy2017}. In the $N\rightarrow\infty$ limit, these lead to $\mathrm{Im}[\Sigma^c_R(0)]\sim 1/\sqrt{T}$ when the $f$'s are described by SYK models. This is more relevant than the MFL self-energy ($\sim T$) at low $T$, but less relevant at high $T$. Thus, once again, if the coefficients of these bilinears are small, then the MFL self-energy will dominate above a certain temperature scale, and the MFL behavior will be restored. 

\section{Boltzmann equation for the marginal-Fermi liquid}
\label{BZE} 

We provide a derivation of (\ref{BE1}). We follow the notation, style, and mechanics of Chapter {\color{red} 5} of Ref.~\onlinecite{Kamenev2011}. The general off-shell Boltzmann equation for modes close to the isotropic Fermi surface ($|p|\approx p_F$; we do not use boldface for momentum-space vectors) is given by
\beq
-\left[\left(i\partial_t + v_F|\nabla+\mathbf{A}_E+\mathbf{A}_B|\right) \circ, F\right] = \Sigma_K^c - (\Sigma_R^c\circ F - F\circ \Sigma_A^c),
\label{BE2}
\eeq
where $F(t,\mathbf{r},p,\omega) = 1 - 2(n_f(\omega)+\delta n(t,\mathbf{r},p,\omega))$ is a parameterization of the distribution function, $\mathbf{A}_E(t)$ and $\mathbf{A}_B(\mathbf{r})$ are parts of the electromagnetic vector potential giving rise to the uniform electric and magnetic fields respectively, with $-d\mathbf{A}_E(t)/dt = \mathbf{E}(t)$ and $\nabla\times\mathbf{A}_B(\mathbf{r}) = \mathcal{B}\hat{z}$ ($\nabla$ denotes the spatial gradient). $\Sigma^c_{R,A,K}$ are the retarded, advanced, and Keldysh components of the conduction electron self-energy respectively. The equation (\ref{BE2}) follows from the Dyson equation for two-point functions on the Keldysh contour~\cite{Kamenev2011}, and hence is exact due to the large $M,N$ limits. The $\circ$ denotes the convolution 
\beq
Z = X\circ Y \Rightarrow Z(t_1,\mathbf{r}_1,t_2,\mathbf{r}_2) = \int dt_3d^2\mathbf{r}_3~X(t_1,\mathbf{r}_1,t_3,\mathbf{r}_3) Y(t_3,\mathbf{r}_3,t_2,\mathbf{r}_2),
\eeq
in the two-coordinate representation, and the $[.~,~.]$ denotes a commutator. We will however mostly use the central-relative coordinate representation instead, with  $p,\omega$ being Fourier transforms of the relative coordinate $\mathbf{r}_1-\mathbf{r}_2,t_1-t_2$, and $\mathbf{r},t$ denoting the central coordinate $(\mathbf{r}_1+\mathbf{r}_2)/2,(t_1+t_2)/2$; this convolution can then be appropriately re-expressed in this representation following Ref.~\onlinecite{Kamenev2011}.

We then use a coordinate remapping $k=p+\mathbf{A}_B(\mathbf{r})$~\cite{Thomas1966,Langreth1966} to redefine $F(t,\mathbf{r},p,\omega) = 1 - 2(n_f(\omega)+\delta n(t,\mathbf{r},p,\omega)) \Rightarrow F(t,k,\omega) = 1 - 2(n_f(\omega)+\delta n(t,k,\omega))$. This is valid as long as the Fermi energy is large enough to make effects of Landau quantization insignificant at the fields in question. The only $\mathbf{r}$ dependence in $F$ then is fictitious, coming from the $\mathbf{r}$ dependence of $\mathbf{A}_B$, and should not affect physical results for spatially uniform transport quantities due to gauge-invariance. It is now absorbed into an implicit $\mathbf{r}$ dependence in $k$. 

We consider the part of (\ref{BE2}) proportional to the infinitesimal $\mathbf{E}(t)$. Because of the isotropy of the Fermi surface and the scattering, we then use the ansatz  $\delta n(t,k,\omega) = k\cdot\varphi(t,\omega)$. We use a first-order gradient expansion in spatial and time derivatives with respect to the central coordinate, which is justified by the spatial uniformity of $\mathbf{E}(t)$ and $B$, and the slow temporal variation of $\mathbf{E}(t)$. The change in the momentum-integrated Keldysh conduction electron Green's function caused by $\mathbf{E}(t)$ through $\delta n$ then is~\cite{Kamenev2011}
\begin{align}
&\delta G^c_K(t,\omega) \equiv \int d^2k~\delta G^c_K(t,k,\omega)=-2\int d^2k\left(G^c_R(|k|,\omega)-G^c_A(|k|,\omega)\right)\delta n(t,k,\omega) \nn
&-2i\int d^2k~\partial_\omega\mathrm{Re}[G^c_R(|k|,\omega)]\partial_t \delta n(t,k,\omega)+2i\int d^2k~\partial_k \mathrm{Re}[G^c_R(|k|,\omega)] \cdot \nabla \mathbf{A}_B(\mathbf{r})\cdot \partial_k \delta n(t,k,\omega) = 0,
\label{GfKchange}
\end{align}
as $G_f^{R,A}$ are isotropic. We have used $\nabla \delta n(t,k,\omega) = \nabla \mathbf{A}_B(\mathbf{r})\cdot\partial_k\delta n(t,k,\omega)$, due to the implicit $\mathbf{r}$ dependence in $k$. The retarded and advanced conduction electron Green's functions are not changed by the applied electric field, as they are only influenced by the change in the distribution $\delta n$ through the self-energies~\cite{Kamenev2011}, which as we show below, are unaffected by the applied electric field. 

On the Keldysh contour, the conduction electron self-energy is given by, analogous to (\ref{Dysonsaddle}),
\beq
\Sigma^c(t_1,t_2) = -g^2G^c(t_1,t_2)G(t_1,t_2)G(t_2,t_1),~~\mathrm{or}~~\Sigma^c_{>,<}(t_1,t_2) = -g^2G^c_{>,<}(t_1,t_2)G_{>,<}(t_1,t_2)G_{<,>}(t_2,t_1).
\eeq
Using the standard relations between the $>,<$ representation and the $R,A,K$ representation~\cite{Kamenev2011,Eberlein2017}, the changes in the conduction electron self-energies due to $\delta n$ are then given by
\begin{align}
&\delta\Sigma^c_R(t_1,t_2)=-\frac{g^2}{4}\theta(t_1-t_2)\delta G^c_K(t_1,t_2) (G_K(t_1,t_2)G_A(t_2,t_1)+G_K(t_2,t_1)G_R(t_1,t_2)), \nn
&\delta\Sigma^c_A(t_1,t_2)=-\frac{g^2}{4}\theta(t_2-t_1)\delta G^c_K(t_1,t_2) (G_K(t_1,t_2)G_R(t_2,t_1)+G_K(t_2,t_1)G_A(t_1,t_2)), \nn
&\delta\Sigma^c_K(t_1,t_2)=-\frac{g^2}{4}\delta G^c_K(t_1,t_2) (G_K(t_1,t_2)G_K(t_2,t_1)+G_R(t_1,t_2)G_A(t_2,t_1)),~~(t_1>t_2), \nn
&\delta\Sigma^c_K(t_1,t_2)=-\frac{g^2}{4}\delta G^c_K(t_1,t_2) (G_K(t_1,t_2)G_K(t_2,t_1)+G_A(t_1,t_2)G_R(t_2,t_1)),~~(t_1<t_2),
\end{align}
which vanish due to (\ref{GfKchange}). Here, $G_{R,A,K}$ denote the island electron Green's functions at equilibrium. Similarly, for the islands, we also get $\delta\Sigma_{R,A,K}=0$, for the same reason. 

The $\mathcal{O}(\mathbf{E})$ part of the RHS of (\ref{BE2}) then is $2(\Sigma^c_R\circ\delta n-\delta n\circ \Sigma^c_A)$.  Using the $p,k,\mathbf{r}$-independence of the by definition $t$-independent equilibrium self-energies $\Sigma^c_{R,A,K}$, and a first-order gradient expansion in central time derivatives, the RHS of (\ref{BE2}) reduces to~\cite{Kamenev2011} 
\beq
4i\mathrm{Im}[\Sigma^c_R(\omega)]\delta n(t,k,\omega) + 2i\partial_\omega\mathrm{Re}[\Sigma^c_R(\omega)]\partial_t\delta n(t,k,\omega).
\label{BERHS}
\eeq

We now turn to the part of the LHS of (\ref{BE2}) proportional to $\mathbf{E}(t)$. Following Sec. {\color{red}5.7} of Ref.~\onlinecite{Kamenev2011}, and noting that the Wigner transform of $\nabla + A_B(\mathbf{r})$ is $k$, it reduces in the first-order gradient expansion in central spatial and time derivatives to
\beq
2i\partial_t\delta n(t,k,\omega)+2i\left(-v_F \partial_t |k+\mathbf{A}_E(t)| n_f^\prime(\omega) +v_F \nabla |k|\cdot \partial_k \delta n(t,k,\omega)-v_F\partial_k |k|\cdot \nabla \mathbf{A}_B(\mathbf{r})\cdot \partial_k \delta n(t,k,\omega)\right),
\label{BELHS}
\eeq
After some algebra, this further reduces to
\beq
2i\partial_t\delta n(t,k,\omega)+2iv_F\mathbf{E}(t)\cdot\hat{k}n_f^\prime(\omega)+ 2iv_F\mathcal{B}(\hat{k}\times\hat{z})\cdot\partial_k\delta n(t,k,\omega).
\eeq
Then, combining this with (\ref{BERHS}), we recover (\ref{BE1}).  The solution to (\ref{BE1}) then shows our ansatz $\delta n(t,k,\omega) = k\cdot\varphi(t,\omega)$ to be self-consistent.

\bibliography{magnet}

\end{document}